%% file: pasa.tex
\documentclass[
  journal=pasa,
  manuscript=research-paper,
  year=2025,
  volume=YY,
]{cup-journal}

\usepackage{amsmath,amsfonts,amssymb}
\usepackage[nopatch]{microtype}
\usepackage{booktabs}
\usepackage{orcidlink}
\usepackage{pdflscape}
\PassOptionsToPackage{breakurl,breaklinks,breaklines}{hyperref}

\usepackage{seqsplit}

\begingroup
\gdef\pkg#1{%
  \texttt{%
    \begingroup
    \catcode46=\active
    \catcode95=\active
    \def.{.\discretionary{-}{}{}}%
    \def_{\_\discretionary{-}{}{}}%
    \seqsplit{#1}%
    \endgroup
  }%
}
\endgroup

\usepackage{xcolor}
\usepackage[normalem]{ulem}

\usepackage{tikz}
\usepackage{tikz}
\usetikzlibrary{shapes.geometric, arrows.meta, positioning}

\tikzset{
    empty/.style={inner sep=0, minimum size=0},
    node distance=2cm,
    process/.style={rectangle, minimum width=2cm, minimum height=1cm, text centered, draw=black},
    database/.style={cylinder, shape border rotate=90, aspect=0.25, draw, text centered},
    decision/.style={diamond, minimum width=2cm, minimum height=1cm, text centered, draw=black},
    arrow/.style={thick,->,>=stealth},
}

\usepackage{epigraph}
\usepackage{longtable}
\usepackage{graphicx}
\usepackage{amsmath}
\usepackage{amssymb}
\usepackage{booktabs}
\usepackage{subcaption}
\usepackage{makecell}
\usepackage[htt]{hyphenat}
\usepackage{xurl}
\definecolor{dodgerblue}{RGB}{30, 144, 255}
\definecolor{crimson}{RGB}{220, 20, 60}
\definecolor{darkerblue}{RGB}{0, 0, 139}
\hypersetup{colorlinks,citecolor=dodgerblue,linkcolor=blue,urlcolor=darkerblue}

\usepackage{siunitx}
\usepackage[nolist,nohyperlinks]{acronym}

\usepackage{listings}
\DeclareSIUnit\degreesq{deg\text{$^2$}}
\DeclareSIUnit\perdegreesq{deg\text{$^{-2}$}}

\newcommand{\degr}{^{\circ}}
\newcommand{\jansky}{\text{Jy}}
\usepackage{svg}
\svgsetup{inkscapeversion=1}
\usepackage{multicol}

\usepackage{listings}

\lstdefinelanguage{YAML}{
  keywords={true,false,null,yes,no},
  keywordstyle=\color{blue}\bfseries,
  sensitive=false,
  comment=[l]{\#},
  morecomment=[l]{\#},
  commentstyle=\color{gray}\ttfamily,
  stringstyle=\color{black},
  moredelim=[s][\bfseries\color{black}]{:}{\ },
  morestring=[b]",
}

\graphicspath{{./}{figures/}}

\title{The Rapid ASKAP Continuum Survey VII: Spectra and Polarisation In Cutouts of Extragalactic Sources (SPICE-RACS) Second Data Release -- Unveiling the Magnetised Sky}

\author{Alec J.M. Thomson \orcidlink{0000-0001-9472-041X}}
\affiliation{SKA Observatory, SKA-Low Science Operations Centre, 26 Dick Perry Avenue, Kensington WA 6151, Australia}
\alsoaffiliation{ATNF, CSIRO Space \& Astronomy, PO Box 1130, Bentley, WA 6102, Australia}
\email[Alec J.M. Thomson]{\url{alec.thomson@skao.int}}

\author{Timothy J. Galvin \orcidlink{0000-0002-2801-766X}}
\affiliation{ATNF, CSIRO Space \& Astronomy, PO Box 1130, Bentley, WA 6102, Australia}
\alsoaffiliation{International Centre for Radio Astronomy Research, Curtin University, Kent St, Bentley WA 6102, Australia}

\author{Stefan W. Duchesne \orcidlink{0000-0002-3846-0315}}
\affiliation{ATNF, CSIRO Space \& Astronomy, PO Box 1130, Bentley, WA 6102, Australia}

\author{Emil Lenc \orcidlink{0000-0002-9994-1593}}
\affiliation{ATNF, CSIRO Space \& Astronomy, PO Box 76, Epping, NSW 1710, Australia}

\author{George Heald \orcidlink{0000-0002-2155-6054}}
\affiliation{SKA Observatory, SKA-Low Science Operations Centre, 26 Dick Perry Avenue, Kensington WA 6151, Australia}
\alsoaffiliation{ATNF, CSIRO Space \& Astronomy, PO Box 1130, Bentley, WA 6102, Australia}

\author{Ondrej Hlinka}
\affiliation{CSIRO Information Management \& Technology, PO Box 883, Kenmore, QLD 4069, Australia}

\author{Sunil Malik \orcidlink{0000-0003-4147-626X}}
\affiliation{Departamento de Física de la Tierra y Astrofísica \& IPARCOS-UCM, Universidad Complutense de Madrid, 28040 Madrid, Spain}

\author{Craig S. Anderson \orcidlink{0000-0002-6243-7879}}
\affiliation{Research School of Astronomy \& Astrophysics, The Australian National University, Canberra ACT 2611, Australia}

\author{Erik Osinga \orcidlink{0000-0002-5815-8965}}
\affiliation{Dunlap Institute for Astronomy and Astrophysics, University of Toronto, 50 St.\ George Street, Toronto, ON M5S 3H4, Canada}

\author{Lerato Baidoo \orcidlink{0000-0003-0520-0696}}
\affiliation{Dunlap Institute for Astronomy and Astrophysics, University of Toronto, 50 St.\ George Street, Toronto, ON M5S 3H4, Canada}

\author{N. M. McClure-Griffiths \orcidlink{0000-0003-2730-957X}}
\affiliation{Research School of Astronomy \& Astrophysics, The Australian National University, Canberra ACT 2611, Australia}
\alsoaffiliation{SKA Observatory, Jodrell Bank, Lower Withington, Macclesfield, SK11 9FT, UK}

\author{Sebastian Hutschenreuter \orcidlink{0000-0002-6952-9688}}
\affiliation{University of Vienna, Department of Astrophysics, T\"urkenschanzstra{\ss}e 17, 1180 Vienna, Austria}

\author{Shane P. O'Sullivan \orcidlink{0000-0002-3968-3051}}
\affiliation{Departamento de Física de la Tierra y Astrofísica \& IPARCOS-UCM, Universidad Complutense de Madrid, 28040 Madrid, Spain}

\author{Takuya Akahori \orcidlink{0000-0001-9399-5331}}
\affiliation{Mizusawa VLBI Observatory, National Astronomical Observatory Japan,
2-21-1 Osawa, Mitaka, Tokyo 181-8588, Japan}

\author{B. M. Gaensler \orcidlink{0000-0002-3382-9558}}
\affiliation{Department of Astronomy and Astrophysics, University of California Santa Cruz, 1156 High Street, Santa Cruz, CA 95064, USA}
\alsoaffiliation{Dunlap Institute for Astronomy and Astrophysics, University of Toronto, 50 St. George Street, Toronto, ON M5S 3H4, Canada}
\alsoaffiliation{David A. Dunlap Department of Astronomy and Astrophysics, University of Toronto, 50 St. George Street, Toronto, ON M5S 3H4, Canada}

\author{J. P. Leahy \orcidlink{0000-0003-2514-9592}}
\affiliation{Jodrell Bank Centre for Astrophysics, Department of Physics and Astronomy, University of Manchester, Manchester M13 9PL, UK}

\author{Y. K. Ma \orcidlink{0000-0003-0742-2006}}
\affiliation{Max-Planck-Institut f\"ur Radioastronomie, Auf dem H\"ugel 69, 53121 Bonn, Germany}

\author{Vanessa A. Moss \orcidlink{0000-0002-3005-9738}}
\affiliation{ATNF, CSIRO Space \& Astronomy, PO Box 76, Epping, NSW 1710, Australia}
\alsoaffiliation{Sydney Institute for Astronomy, School of Physics A28, University of Sydney, NSW 2006, Australia}

\author{L. Rudnick \orcidlink{0000-0001-5636-7213}}
\affiliation{Minnesota Institute for Astrophysics, University of Minnesota, 116 Church Street SE, Minneapolis, MN 55455, USA}

\author{C. L. Van Eck \orcidlink{0000-0002-7641-9946}}
\affiliation{Research School of Astronomy \& Astrophysics, The Australian National University, Canberra ACT 2611, Australia}
\alsoaffiliation{Dunlap Institute for Astronomy and Astrophysics, University of Toronto, 50 St. George Street, Toronto, ON M5S 3H4, Canada}

\author{J. L. West \orcidlink{0000-0001-7722-8458}}
\affiliation{National Research Council Canada, Herzberg Research Centre for Astronomy and Astrophysics, Dominion Radio Astrophysical Observatory, PO Box 248,
Penticton, BC V2A 6J9, Canada}
\alsoaffiliation{Dunlap Institute for Astronomy and Astrophysics, University of Toronto, 50 St. George Street, Toronto, ON M5S 3H4, Canada}


\begin{document}

\input{content}

\printendnotes

\bibliography{correct}

\appendix

\input{appendix}

\end{document}

%% file: content.tex
\begin{abstract}
We present the second data release (DR2) of Spectra and Polarisation in Cutouts of Extragalactic sources from RACS (SPICE-RACS). SPICE-RACS DR2 is derived from the third low-band epoch of the Rapid ASKAP Continuum Survey (RACS-low3) and covers the entire sky from the South celestial pole up to a declination of $+\ang{49}$; approximately \qty{87.5}{\percent} of the celestial sphere. We produce `cutout' spectral cubes in Stokes $I$, $Q$, $U$ around 4\,million radio sources and extract spectra towards 5\,million radio components. Across our observed band of \qtyrange{799.5}{1087.5}{\mega\hertz} we find an $rms$ noise of $\sim\qty{200}{\micro\jansky\per PSF}$, an angular resolution of $\sim\ang{;;15}$, and residual wide-field instrumental polarisation on the order of $\qty{0.1}{\percent}$. After de-duplication, our polarisation catalogue contains the detection of $2.5\times10^5$ ($3.4\times10^5$) Faraday rotation measures (RM) for components with a linearly polarised signal above $8\sigma$ ($6\sigma$). This places SPICE-RACS DR2 as the largest single RM catalogue ever produced by nearly an order of magnitude; the number of RMs in our catalogue alone is $\sim5$ times larger than every previous RM catalogue combined. Our resulting RM grid has an areal density of $6.7 \substack{+1.8\\-1.7}\,\unit{\perdegreesq}$, providing an effective `resolution' of $\sim\ang{;23}$, and reveals striking features across the sky. The broad-band RMs have a median uncertainty of $\sim$\qty{2}{\radian\per\metre\squared}, and include complexity metrics and information from the time domain. The breadth and quality of the SPICE-RACS DR2 dataset will enable a new generation of RM science. Further, SPICE-RACS will provide an ideal reference for forthcoming deep polarisation surveys such as the ASKAP POSSUM survey. All of our data products are publicly available on the CSIRO Data Access Portal (DAP) and the CSIRO ASKAP Science Data Archive (CASDA).
\end{abstract}


\section{Introduction}\label{sec:intro}

\defcitealias{McConnell2020}{Paper~I}
\defcitealias{Hale2021}{Paper~II}
\defcitealias{Thomson2023}{Paper~III}
\defcitealias{Duchesne2023}{Paper~IV}
\defcitealias{Duchesne2024}{Paper~V}
\defcitealias{Duchesne2025}{Paper~VI}
\defcitealias{Taylor2009}{NVSS}
\defcitealias{Schnitzeler2019}{S-PASS/ATCA}
\defcitealias{Perley2013}{PB13}
\defcitealias{Perley2017}{PB17}
\defcitealias{Taylor2024}{TL24}
\defcitealias{VanEck2023}{RMTable}
\defcitealias{Manchester2005}{PSRCAT}


The Faraday effect provides a unique probe of the magneto-ionic medium of the universe. The Faraday rotation measure (RM) of a linearly polarised radio source is proportional to the product of the thermal electron density and the magnetic field parallel to and integrated along the line of sight. Through collecting large numbers of RMs across the sky from background sources, we can execute a `rotation measure grid' experiment~\citep{Gaensler2004}. These RM grids can then be used to illuminate foreground objects of interest such as the Milky Way~\citep[e.g.][]{Dickey2022}, nearby galaxies~\citep[e.g.][]{Livingston2022, Livingston2024, Stein2025}, groups and clusters~\citep[e.g.][]{Anderson2021, Anderson2024, Osinga2022}, and even the cosmic web~\citep[e.g.][]{Carretti2022,Carretti2023,Carretti2025}. Using an external constraint on electron density, such as H$\alpha$ emission, it is possible to use RM grids to measure magnetic field strength and structure \citep[see e.g.][]{Harvey-Smith2011a}. Conversely, the line-of-sight magnetic fields can act as a `dye tracer', allowing RM grids to provide a uniquely sensitive probe of low-density thermal gas~\citep{Anderson2021,Anderson2024}.

The scientific yield of an RM grid is determined by four primary observational factors. First, we require radio continuum observations across a broad bandwidth. At its most simplistic, sampling many values of wavelength-squared ($\lambda^2$) circumvents ambiguities in the determined RM. \citet{Brentjens2005} and \citet{Dickey2018} further describe the impacts of channelisation on the RM spread function (RMSF). \citet{Brentjens2005} give the full-width at half-maximum (FWHM) Faraday depth resolution ($\delta\phi$), the `maximum Faraday depth scale' ($\phi_\text{max-scale}$), and the maximum Faraday depth ($\phi_\text{max}$) in their Equations 61, 62, and 63 respectively. We note that $\phi_\text{max}$ is inversely proportional to the nominal channel size in $\lambda^2$ ($\delta\lambda^2$). Having sufficiently large $\phi_\text{max-scale}$ and with respect to $\delta\phi$ allows the `Faraday complexity' to be resolved in Farday depth~\citep[see e.g.][]{Alger2021}. Second, we require a survey across a wide area of the sky. Wide areal coverage is needed to characterise large-scale foreground structure, such as the RM signature of the Milky Way. Third, we desire a survey with high sensitivity. The areal density of RMs across the sky determines the `resolution' of our RM grid. Finally, we require high angular resolution. Such resolution is required to enable the study of the intrinsic individual sources, such as the lobes of radio galaxies~\citep[e.g.][]{Anderson2018,Anderson2018a} or from the diffuse emission of the Milky Way~\citep[e.g.][]{Landecker2012, Erceg2024}. As discussed in \citet[][hereafter {\citetalias{Thomson2023}}]{Thomson2023}, the historical compilation of RMs by \citet{VanEck2023} is dominated by the catalogue from the NRAO VLA Sky Survey \citep[][hereafter \citetalias{Taylor2009}]{Condon1998,Taylor2009}. \citetalias{Taylor2009} was derived from two narrowly spaced frequencies at $\sim\qty{1.4}{\giga\hertz}$, and only provides an areal RM density of $\sim\qty{1}{\perdegreesq}$. The Southern hemisphere is more sparsely sampled, with \citet[hereafter {\citetalias{Schnitzeler2019}}]{Schnitzeler2019} only providing an RM density of $\sim\qty{0.2}{\perdegreesq}$ up to a declination of $\ang{0}$.

As a purpose-built survey instrument, the Australian SKA Pathfinder radio telescope~\citep[ASKAP,][]{Hotan2021} enables a generational leap in polarisation surveys. The primary polarisation Survey Science Project with ASKAP is Polarisation Sky Survey of the Universe's Magnetism~\citep[POSSUM,][]{Gaensler2025}. At the end of the 5-year survey campaigns POSSUM is expected to provide $\sim7.75\times10^5$ ($\sim8.77\times10^5$) RMs covering $\qty{2\pi}{\steradian}$ ($\qty{1.47\pi}{\steradian}$) at \qtyrange{800}{1088}{\mega\hertz} (\qtyrange{800}{1088}{\mega\hertz} + \qtyrange{1296}{1440}{\mega\hertz}) \citep{Vanderwoude2024}. We note that the low-frequency component of POSSUM is fully commensal with the Evolutionary Map of the Universe survey~\citep[EMU,][]{Hopkins2025}.

The ASKAP Observatory has conducted the Rapid ASKAP Continuum Survey~\citep[RACS,][hereafter \citetalias{McConnell2020}]{McConnell2020} both in preparation for, and in support of, the full Survey Science Projects such as POSSUM. RACS is being conducted over multiple epochs and covers the full frequency range available to ASKAP. To date 5 epochs of RACS have been observed, namely; RACS-low1~\citepalias{McConnell2020} and low2 at \qty{887.5}{\mega\hertz}, low3 at \qty{943.5}{\mega\hertz} (this work, and Galvin et al. in prep), mid at \qty{1367.5}{\mega\hertz}~\citep[][hereafter {\citetalias{Duchesne2023}}]{Duchesne2023}, and high at \qty{1655.5}{\mega\hertz}~\citep[][hereafter {\citetalias{Duchesne2025}}]{Duchesne2025}. Catalogues in total intensity were presented by \citet[][hereafter {\citetalias{Hale2021}}]{Hale2021} for RACS-low1, \citet[][hereafter {\citetalias{Duchesne2024}}]{Duchesne2024} for RACS-mid, and \citetalias{Duchesne2025} for RACS-high.

In \citetalias{Thomson2023} we described the first data release (DR1) of Spectra and Polarisation in Cutouts of Extragalactic sources from RACS (SPICE-RACS). SPICE-RACS is a joint project between the ASKAP Observatory and the POSSUM Collaboration. Its aim is to derive linear polarisation results from RACS using spatial cutouts around sources detected in total intensity. The DR1 catalogue was derived from 30 RACS-low1 fields with a sensitivity of $\sim\qty{150}{\micro\jansky\per PSF}$ in linear polarisation~\citep{Thomson2024}. The catalogue provided detections of 5818 linearly polarised components over $\sim\qty{1300}{\degreesq}$; an areal RM density of $\sim\qty{4.5}{\perdegreesq}$.

In this paper, we introduce the second SPICE-RACS data release (DR2), which covers the entire Southern sky visible to ASKAP; a total area of $\qty{3.5\pi}{\steradian}$. This paper is structured as follows. In \S\ref{sec:data}, we describe the observations and data used in this data release. In \S\ref{sec:processing}, we explain our data processing and pipeline. Next, in \S\ref{sec:catalogue}, we provide an overview of the catalogues included in this paper. We then present a brief analysis of the data in \S\ref{sec:analysis}, and set out our conclusions in \S\ref{sec:conclusion}.

\section{Data}\label{sec:data}

\begin{figure}
    \centering
    \includegraphics[width=\linewidth]{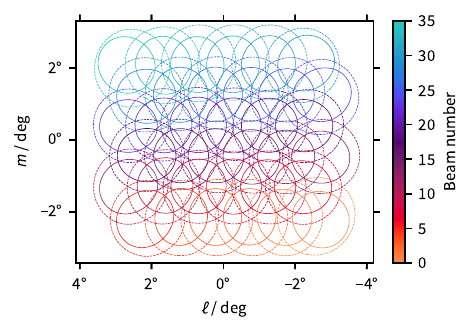}
    \caption{
        The field of view of RACS-low3 as measured by holography. We show the 50\% level of the Stokes I response beams at \qty{800}{\mega\hertz} and \qty{1088}{\mega\hertz} in solid and dashed contours, respectively. We show the beam positions with respect to the telescope pointing centre, and colour each beam by its respective number.
    }
    \label{fig:footprint}
\end{figure}

For this second data release we elect to use the third epoch of the low-band RACS (RACS-low3) as it provides a number of improvements over previous epochs. RACS-low3 uses a central frequency of \qty{943.5}{\mega\hertz} (matching the central frequency of EMU/POSSUM), which in turn yields a lower system temperature \citep{Hotan2021}. Unlike RACS-low1 and low2, RACS-low3 was observed with a \verb|closepack_36| footprint, with a sky tiling exactly matching RACS-mid and RACS-high \citepalias[see][]{Duchesne2023}, as well as the EMU/POSSUM survey. We show the footprint of the RACS-low3 beams in Figure~\ref{fig:footprint}. The ASKAP Observatory conducted an initial observing run of RACS-low3 from December 2023 to February 2024, with an additional set of re-observations completed from May to April 2024. The re-observations were requested to help improve a small subset of data that had reduced data quality and high levels of flagging. Here we will use the initial set of observations, and leave the inclusion of the later re-observations to a future data release.

The initial observing campaign for RACS-low3 consisted of 1615 target observations of 1493 unique fields. Each observation is uniquely identified by a `scheduling block identifier' (SBID), with each field having its own unique name derived from its central coordinates. RACS-low3 was observed under a single set of `parent' beamforming weights\footnote{Here `parent' refers to the set of weights derived from an initial observation of the Sun. Subsequent weights are produced using the on-dish calibrator system with the purpose of keeping the system response in line with the initial `parent' weights. See \citet{Hotan2021} for further details.} (with SBID 55217), meaning that the formed beams should be common amongst all observations. The ASKAP Observatory conducted holography of the RACS-low3 beams at the start and end of the observing run with SBIDs 55219 and 59276, respectively. Throughout this entire work we use the holography from SBID 55219 for self-consistency.

Each individual RACS-low3 observation was processed autonomously by the ASKAP Observatory pipeline \citep{Guzman2019}. This processing included flagging, primary bandpass calibration, and total intensity imaging with self-calibration. For further details on this processing pipeline, we refer the reader to \citetalias{McConnell2020}. The resulting total intensity images cover the entire Southern sky up to $\delta\sim+\ang{49}$ with a median interquartile range (IQR)\footnote{Throughout this work we will show the IQR resulting from the $16\text{th}$ and $84\text{th}$ percentiles.} $rms$ noise of $190 \substack{+60\\-32}\,\si{\micro\jansky\per PSF}$. The calibrated visibilities, mosaicked images, and initial source lists for each field were deposited and released on the CSIRO ASKAP Data Archive \citep[CASDA,][]{Huynh2020}.

For this data release, we will present the polarisation results from processing each individual RACS-low3 field independently. Whilst the edges of each field will lack the maximum possible sensitivity, the time-domain information will be retained. This data release will therefore be similar to the `time domain' catalogue from \citetalias{Duchesne2024}. We will publish a `full sensitivity' catalogue in a future data release.

\section{Processing}\label{sec:processing}

\subsection{Initial cataloguing}

For the purpose of guiding polarised source extraction, we make use of a total intensity source catalogue constructed from the Observatory-processed RACS-low3 Stokes $I$ images. The construction of the Stokes $I$ catalogue follows processes used for previous RACS epochs (\citetalias{Hale2021,Duchesne2024,Duchesne2025}) and makes use of the overlapping images to form full-sensitivity mosaics for each unique RACS-low3 field. Unlike our polarisation catalogue, this total intensity source-finding catalogue is de-duplicated.

The PSF of the RACS-low3 observations varies significantly over the survey region, largely as a function of declination, with a major axis FWHM ranging from $\sim\qtyrange{12}{60}{arcsec}$. To account for this variation, we follow \citetalias{Duchesne2024} and use the lowest common resolution from all overlapping images that make up each mosaic rather than a fixed angular resolution over the whole survey. We use \texttt{beamcon\_2D} from RACS-tools \footnote{\url{https://github.com/AlecThomson/RACS-tools}} to find and convolve to the smallest fitting PSF. Once the individual images are convolved for each mosaic, we use \textsc{SWarp} \citep{swarp} for regridding and co-addition, weighting individual images by the output weight files from the \textsc{ASKAPsoft} processing. These weights are a combination of the primary beam attenuation and the number of unflagged visibilities for a given PAF beam dataset.

Following \citetalias{Duchesne2024}, we use \textsc{PyBDSF} \footnote{\url{https://github.com/lofar-astron/PyBDSF}, v1.10.3.} \citep{pybdsf} for source-finding and characterisation. The source lists for each mosaic are then merged, with duplicate measurements of sources in overlapping regions removed to retain only a single measurement for each source. \textsc{PyBDSF} produces a list of 2D Gaussian components alongside each source list. We have found previously that sources can be modelled with more Gaussian components than necessary \citepalias[][]{Duchesne2024}, so for sources that have a ratio of integrated to peak flux density less than the median for a given mosaic source list, we assume these are true point sources and regroup their components (if more than one component) into a single component. This is done by summing the flux density and taking the size as reported in the corresponding source list. The position is also taken from the source list rather than an average from the component list. Source lists for each mosaic are then merged together, with duplicate measurements in overlapping regions removed following \citetalias{Duchesne2024}. The contiguous, de-duplicated catalogues contain 3\,980\,923 sources and 5\,479\,676 Gaussian components.

\subsection{Arrakis}\label{sec:arrakis}
In \citetalias{Thomson2023} we presented our highly parallelised processing pipeline, \textsc{Arrakis}\footnote{\url{https://github.com/alecThomson/arrakis}}, which was used to derive the DR1 catalogue. \textsc{Arrakis} is built in Python using the \textsc{Prefect} and \textsc{Dask}~\citep{dask} frameworks; allowing it to be run and scaled on a huge number of hardware platforms. Version 1 (v1) of \textsc{Arrakis} was a post-processing pipeline which required us to use portions of the RACS and ASKAP pipelines to produce images for polarisation analysis. Here we present v2 of \textsc{Arrakis}, which is a fully self-contained pipeline that can take calibrated ASKAP visibilities, such as those available on CASDA, and produce polarisation images and catalogues ready for scientific use. Here we will provide a description of the new, or substantially modified, components of \textsc{Arrakis}. Unless otherwise described, the primary functions of \textsc{Arrakis} remains the same as in \citetalias{Thomson2023}. All of our \textsc{Arrakis} processing was run on the CSIRO High Performance Computing cluster \textit{Petrichor}.

\subsubsection{Spectro-polarimetric imaging}\label{sec:imaging}

In producing images for polarisation analysis from ASKAP we have two primary requirements. First, we require suitable handling of the non-co-planar effects \citep[i.e. the `$w$-term',][]{Cornwell2008} when producing wide-field images. Second, and perhaps most critically, we require multiple output channel images in order to perform rotation measure synthesis~\citep[RM-synthesis,][]{Brentjens2005}. With these requirements in mind, we chose to use the \textsc{WSClean} imager~\citep{offringa2014,offringa2017}. The \pkg{wgridder} algorithm~\citep{Arras2020, Ye2022} in \textsc{WSClean} provides highly efficient handling of the $w$-term, and is parallelisable across multiple channels images when gridding and de-gridding visibilities. Further, when producing a frequency cube across multiple Stokes parameters the options \texttt{join-channels}, \texttt{join-polarizations}, and \texttt{squared-channel-joining} options enable \textsc{WSClean} to clean the narrow channels utilising the information from the full bandwidth and polarisations. This provides improved image quality in comparison to the Taylor-term imaging or per-channel imaging performed by the \textsc{YandaSoft} \pkg{cimager}/\pkg{imager}~\citep{Guzman2019} used for SPICE-RACS DR1. In \textsc{Arrakis} v2 we have implemented a thin Python wrapper for \textsc{WSClean} which is called within the \pkg{arrakis.imager} pipeflow flow. 

Visibilities produced and calibrated by the \textsc{ASKAPsoft} pipeline are not directly compatible with imagers like \textsc{WSClean}. First, the reported fields of each of the 36 beams are reported as the centre of the field, which is not the same as the pointing centre or phase centre of any of the individual beams. A common field centre would be needed in the case of joint-beam deconvolution, however we chose to image each beam independently. Furthermore, the correlations stored in Observatory-processed visibilities are initially in the instrumental frame. ASKAP has a unique `roll' axis which allows the feeds to be rotated arbitrarily with respect to the sky and as a function of time. In practice, however, the roll axis is driven such that ASKAP operates as an equatorial telescope, with a constant rotational offset from the celestial coordinate frame. Finally, \textsc{ASKAPsoft} adopts a Stokes definition in which $I = XX + YY$. \textsc{WSClean}, however, assumes that the input data is in the IAU polarisation standard, where $X$ and $Y$ align with celestial North and East respectively. \textsc{WSClean} also adopts the Stokes convention that $I = (XX + YY) / 2$. We have implemented a Python library \textsc{FixMS}\footnote{\url{https://github.com/AlecThomson/FixMS}} which can convert \textsc{ASKAPsoft} visibilities into a format compatible with \textsc{WSClean}. We detail the formalism of this conversion in \ref{sec:vis_convert}. For SPICE-RACS DR2 we use version v0.2.9 of \textsc{FixMS}.

We make use of a common set of \textsc{WSClean} options, with some direction-dependent settings. For all images we enable the \pkg{wgridder} and \pkg{join-channels} options. We image Stokes $I$ independently, and Stokes $Q$ and $U$ jointly with \texttt{join-polarizations} and \texttt{squared-channel-joining}. By default we use an image size of 6144 pixels, with a cell size of \ang{;;2.5}, and a visibility weighting of Briggs robust $-0.5$~\citep{Briggs1995}. We apply a cut in $uv$-space of 200 wavelengths to suppress any short-baseline interference (such as the Sun). In the case of \textsc{clean} divergence, usually caused by a bright source outside the image boundary, we manually increased the image size to encompass the source. We use \textsc{clean} major cycles with a \texttt{mgain} of 0.6. Since our focus for SPICE-RACS is primarily on characterisation of compact components, and we are using a $uv$-cut, we do not enable multiscale cleaning. 

To derive the polarisation properties of our components we use RM-synthesis~\citet{Brentjens2005}, which produces a Faraday depth ($\phi$) dispersion function (FDF) via a Fourier transform. The Faraday depth defines the amount of Faraday rotation integrated along the line of sight. In the most simplistic case the Faraday depth reduces to the RM. Here we will report the Faraday depth at the peak in the FDF as an RM. 

Here we use the software package \textsc{RM-tools}~\citep[][Van Eck in prep.]{Purcell2020} to compute Faraday depth information. \textsc{RM-tools} takes $\delta\lambda^2$ to be the largest channel in $\lambda^2$. We set the number of channels for a given field based on its central Galactic latitude. If a field centre has Galactic latitude $|b|>\ang{10}$ we produce 36 channel images, otherwise we produce 72. These channelisations provide maximum Faraday depths of \qty{\sim630}{\radian\per\square\meter} and \qty{\sim1250}{\radian\per\square\meter}, respectively. This allows us to remain sensitive to the higher magnitude RMs we expect towards the Galactic plane. We summarise these values, and the other properties of the RMSF, in Table~\ref{tab:properties}.

During deconvolution we configured \textsc{WSClean} to derive a direction-dependent measure of the signal-to-noise throughout the image at the start of each major cycle. Subsequently these signal-to-noise measures are used to constrain the deconvolution process to pixels of high significance. Although this largely eliminates issues around over-cleaning (e.g. cleaning towards directions that do not contain genuine sky emission) we found that in some instances the derived mask is too conservative, ultimately making noise-based stopping criteria ineffective at terminating the deconvolution process. We found this to happen towards directions with bright ($\gtrsim\qty{1}{\jansky\per PSF}$) and extended source ($\gtrsim\ang{;10}$) structure. We speculate that the internally derived clean mask does not completely capture all sources in the field as \textsc{WSClean} enters its deep-cleaning mode. Consequently, sidelobe noise from uncleaned sources prevent the residual image noise characteristics from triggering \textsc{WSClean}'s stopping criteria. To counter this behaviour we modified \textsc{WSClean}\footnote{\url{https://github.com/tjgalvin/wsclean}, commit \texttt{83f6743}} to accept an additional parameter to force the derivation of a new signal-to-noise clean mask across a specified set of major cleaning cycles. We set this parameter to 7 while imaging all visibility data in Stokes $I$.

\begin{figure}
    \centering
    \includegraphics[width=\textwidth]{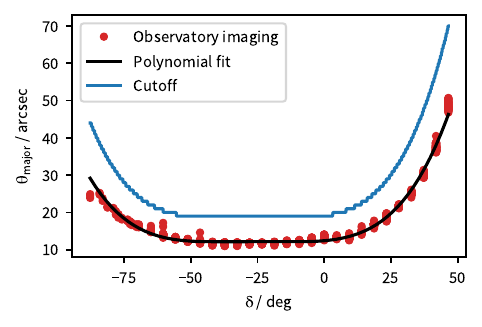}
    \caption{Our cutoff criteria, as based on major axis of the point-spread function (PSF, $\theta$) as a function of Declination ($\delta$). In red we show the $\theta_\text{major}$ derived from the ASKAP Observatory processing of RACS-low3, which used a visibiltiy weighting of Robust 0. In black we show a fitted polynomial with the function form: $\theta_\text{major}=1.2\times10^{1} + 6.3\times10^{-2}\delta + 4.9\times10^{-3}\delta^2 + 1.4\times10^{-4}\delta^3 + 1.3\times10^{-6}\delta^4$. In blue we show the cutoff criteria we have adopted for SPICE-RACS DR2. The cutoff is simply the rounded value (in arcseconds) of the polynomial fit increased by 50\%.}
    \label{fig:psf_dec}
\end{figure}

Since \textsc{WSClean} produces channel images, we form FITS cubes using another simple library we have implemented, \textsc{fits-cube}. We first convolve the channel images to the lowest common resolution for each beam across all channels using \pkg{beamcon\_2D} from \textsc{RACS-tools}~\footnote{\url{https://github.com/AlecThomson/RACS-tools}, v3.0.8}. We set a cutoff to the major axis of the synthesised point-spread function (PSF) as a function of declination. This cutoff is not intended to be our final resolution, rather just a method to flag out any channels with a particularly poor PSF. In Figure~\ref{fig:psf_dec} we show the cut-off criteria we have adopted for SPICE-RACS DR2. We base our cutoff on the initial Observatory processing of RACS-low3, which used a visibility weighting of Briggs robust 0~\citep{Briggs1995}. We fit a polynomial to the main axis of the PSF as a function of declination, and then increase this value by 50\%, rounded up to the nearest arcsecond. Using purely wavelength-dependent scaling, the PSF at the lowest frequency in RACS-low3 should only be $\sim20\%$ higher than the midband value. This fact, in combination with our more uniform weighting, means that our 50\% boost to the PSF cut is simply a coarse cut to remove any particularly poorly behaved channels. After our convolution, we combine each channel image with a single FITS cube per beam and Stokes parameter. These images then flow into our cutout and post-processing procedure.

\subsubsection{Cutouts and post-processing}\label{sec:cutout}
Our post-processing procedure remains mostly the same as described in \citetalias{Thomson2023} (see \S3.3 and \S3.4 of that work for reference). The notable changes for this data release are as follows. First, as described above, we do not impose a global common PSF (we used \ang{;;25} in DR1). Instead, we only convolve each cubelet to the lowest common resolution required for mosaicking. In this way, our survey has a direction-dependent spatial resolution. Second, we have improved our methods for extracting noise spectra. To accommodate for the varying PSF, we extract pixels around each source in an annulus with an inner and outer radius of 1.5 and $5\times$ the minor axis of the cutout PSF, respectively. Further, we now extract the background and noise spectrum simultaneously by fitting a normal distribution to the extracted pixels. Before running any analysis on the spectra (e.g. RM-synthesis) we subtract the background spectra from the flux density spectra for each Stokes parameter. As shown in \citetalias{Thomson2023} (see Appendix B of that work) this annulus method we use for background and noise estimation is robust to overlapping sources near to the component of interest.

\begin{table}
	\caption{Observational properties of SPICE-RACS DR2.}\label{tab:properties}
	\begin{tabular}{lccc}
		\toprule
		\multicolumn{1}{l}{Property} & \multicolumn{2}{c}{Value} & Notes\\
		\midrule
		Shortest baseline / $\lambda$ & \multicolumn{2}{c}{$200$} \\
        Shortest baseline / \si{\m}& \multicolumn{2}{c}{$55.1$ --- $75.0$} & a \\
		Longest baseline / \si{\km} & \multicolumn{2}{c}{$6.4$} \\
		Angular resolution / arcsec & \multicolumn{2}{c}{$11.8$ --- $75.9$}  \\
		J2000 declination / deg & \multicolumn{2}{c}{$-90$ --- $+49.4$}\\
		J2000 right ascension / deg & \multicolumn{2}{c}{$0$ --- $360$} \\
		Areal coverage / \unit{\steradian} & \multicolumn{2}{c}{$3.5\pi$}\\
		Bandwidth / \unit{\mega\hertz} & \multicolumn{2}{c}{$799.5$ --- $1087.5$}\\
        $\lambda^2$ coverage / \unit{\metre\squared} & \multicolumn{2}{c}{$0.076$ --- $0.14$}\\
        $\lambda_0^2$ / \unit{\centi\metre\squared} & \multicolumn{2}{c}{$1010 \substack{+41\\-20}$} & b \\
		Stokes $I$ \textit{rms} noise / \unit{\micro\jansky\per PSF} & \multicolumn{2}{c}{$211 \substack{+150\\-48}$} & b\\
		Stokes $Q$, $U$ \textit{rms} noise / \unit{\micro\jansky\per PSF} & \multicolumn{2}{c}{$194 \substack{+130\\-38}$} & b\\
        $\delta\phi$ FWHM / \unit{rad\,m^{-2}} & \multicolumn{2}{c}{$63.2 \substack{+3.4\\-3.3}$} & b \\
        $\phi_\text{max-scale}$ / \unit{rad\,m^{-2}} & \multicolumn{2}{c}{$41$} & c \\
        \hline
            &  $|b|<\ang{10}$ & $|b|>\ang{10}$ \\
        \hline
		Channel width / \unit{\mega\hertz} & $4$ & $8$\\
        Number of channels & $72$ & $36$ \\
		$\lambda^2$ channel width / \unit{\centi\metre\squared} & $5.7$ --- $14$ & $12$ --- $27$\\
		$\phi_\text{max}$ / \unit{rad\,m^{-2}} & $1250$ & $630$ & c \\
		\bottomrule
	\end{tabular}
    \medskip
    \begin{tablenotes}
        \item[a] Wavelength-dependent
        \item[b] Derived from the final catalogue and reported as the median $\pm$ the 16th and 84th percentiles.
        \item[c] From the uniformly-weighted RMSF assuming no missing channels. 
    \end{tablenotes}
\end{table}

To select our final set of fields for use in this data release we manually inspected the quality of each observed field. Members of the POSSUM collaboration were assigned a number fields and assessed the noise properties and RM quality individually. Of the 1615 observed fields, we identified 31 of reduced data quality. Of this subset, 19 had a repeated observation of superior data quality, so we substituted these into our dataset. The remaining 11 fields are included in this data release, however, we expect that we should be able to replace and improve these fields when we make use of re-observations in a future release. In all, for this data release we make use of 1596 observations of 1493 unique fields. We summarise the key properties of our observations for this data release in Table~\ref{tab:properties}.

\section{Catalogues}\label{sec:catalogue}
We produce our polarisation component catalogue following the \textsc{RMTable} schema~\citep{VanEck2023}. This schema specifies a minimum set of columns to which a user can make additions. We supply the same set of additional columns as in \citetalias{Thomson2023} as well as the following set of new columns:
\begin{itemize}
    \item \verb|l_tile_centre| - Separation from the tile centre in RA.
    \item \verb|m_tile_centre| - Separation from the tile centre in Dec.
    \item \verb|stokesI_bkg| - Band-averaged background flux density in vicinity of the component in Stokes $I$.
    \item \verb|stokesQ_bkg| - As above in Stokes $Q$.
    \item \verb|stokesU_bkg| - As above in Stokes $U$.
    \item \verb|goodI_flag| - A boolean flag that is `True' when the component is included in the our `\texttt{goodI}' subset. See \S\ref{sec:subsets} below for the definitions.
    \item \verb|goodRM_flag| - As above for the `\texttt{goodRM}' subset.
    \item \verb|nn_rm_count| - The number count of nearest neighbours in the \texttt{goodRM} subset used for foreground RM estimation (see \S\ref{sec:nn_foreground}).
    \item \verb|nn_rm_med| - The median RM in nearest neighbour ensemble.
    \item \verb|nn_rm_mean|- The mean RM in nearest neighbour ensemble.
    \item \verb|nn_rm_mad_std| - The median absolute deviation scaled to standard deviation of RM in nearest neighbour ensemble.
    \item \verb|nn_rm_std| - The standard deviation RM in nearest neighbour ensemble.
    \item \verb|nn_rm_se| - The standard error on the mean RM in nearest neighbour ensemble.
    \item \verb|nn_rm_wmean| - The error-weighted mean RM in nearest neighbour ensemble.
    \item \verb|nn_rm_wmean_err| - The standard error on the error-weighted mean RM in nearest neighbour ensemble.
    \item \verb|nn_rm_sep_max_deg| - The largest angular separation in nearest neighbour ensemble.
    \item \verb|nn_rm_sep_min_deg| - The smallest angular separation in nearest neighbour ensemble.
\end{itemize}
We describe the nearest-neighbour statistics in \S\ref{sec:nn_foreground}. For a detailed description of each of the other columns, we refer the reader to \citetalias{Thomson2023}. We provide data access information, including the catalogue, in \S\ref{sec:data_access}.

\subsection{Basic subsets and de-duplication}\label{sec:subsets}

Following \citetalias{Thomson2023}, our catalogue contains a row for every source detected in total intensity. We must therefore stress that not every reported row will correspond to a component with a reliable detection of polarised flux. Rather, we have retained all possible values so that users in the community can make their own cuts on the catalogue to satisfy their own needs and balance of completeness vs. correctness. Better still, following the arguments of \citet{Rudnick2019}, we encourage users of the catalogue data to weight, rather than cut, components in their RM grid.

To assist users of the catalogue, we have again defined some basic subsets that should be reasonable, although conservative, for most science cases. These are as follows:
\begin{itemize}
    \item \texttt{goodI}: Where the \verb|stokesI_fit_flag| and \verb|channel_flag| columns are \verb|False|.
    \item \texttt{goodRM}: \texttt{goodI}, and where \verb|leakage_flag| and \verb|snr_flag| are \verb|False|.
\end{itemize}
The \texttt{goodI} subset will select components where our Stokes $I$ model fit was succesful, and where more than half the bandwidth was intact. The \texttt{goodRM} subset will further select components where the polarised signal-to-noise is above $8\sigma$ and where the fractional polarisation is above our estimated residual leakage (see \S\ref{sec:leakage}).

Since we have produced a time-domain catalogue, we have repeated observations of components from both overlapping field edges and repeated observations. For the purposes of this work we also produce a de-duplicated catalogue by simply selecting duplicated components which lie closest to their respective field centre. Hereafter we refer to our catalogue \textit{including} duplicate components as our `full catalogue', and the other as the `de-duplicated catalogue'. We give the number count of components in our catalogue for our basic subsets with and without de-duplication in Table~\ref{tab:counts}. Note that, as in \citetalias{Thomson2023}, we do not include components in our polarisation catalogue when we cannot successfully fit a model to its Stokes $I$ spectrum. As such a relatively small number of components (415) which are present in the total intensity source-finding catalogues are not included in our polarisation catalogue.

\begin{table}
    \centering
    \begin{tabular}{lccc}
    \toprule
    Subset & Total count & De-duplicated count & Repeated count\\
    \midrule
    All    & 9\,294\,225 & 5\,479\,261 & 3\,814\,964 \\
    \texttt{goodI} & 4\,543\,523 & 3\,122\,559 & 1\,420\,964\\
    \texttt{goodRM} & 333\,174 & 246\,509 & 86\,665\\
    \midrule
    Not blended &  &  & \\
    \midrule
    All    & 7\,623\,061 & 4\,510\,007 & 3\,113\,054 \\
    \texttt{goodI} & 3\,125\,653 & 2\,242\,721 & 882\,932\\
    \texttt{goodRM} & 74\,276 & 58\,372 & 15\,904\\
    \bottomrule
    \end{tabular}
    \caption{
        The number count of Gaussian components in our catalogue having applied a subset or de-duplication. The number of repeated observations is simply the total minus the de-duplicated count. In the lower portion of the table, we give the same set of counts but with the additional constraint that the \texttt{is\_blended\_flag} column is false.
    }
    \label{tab:counts}
\end{table}

As discussed in \citetalias{Thomson2023}, pixel-wise extraction of spectra means that neighbouring components are not statistically separated as they would be by a true source-finding algorithm. As there are certain scientific use-cases that require isolated components and spectra, we again provide the \verb|is_blended_flag| column. This flag will be true if any neighbouring components are within 1 PSF width of the given component. In the lower portion of Table~\ref{tab:counts}, we also show the component counts having selected only isolated components where \verb|is_blended_flag| is false. 

We see that the majority of polarised components (187\,137 deduplicated components, or \qty{76}{\percent}) have been flagged as blended. Such components must therefore belong to a PyBDSF source and are therefore extended. In Table~\ref{tab:source_counts} we show the number counts of PyBDSF sources for the same subsets as Table~\ref{tab:counts}. From this we can see that the 187\,137 deduplicated and blended components in the  \texttt{goodRM} subset corresponds to 79\,829 sources (or $\sim$\qty{60}{\percent}). By comparison, only 181\,004, or $\sim\qty{4.5}{\percent}$, of the sources in the overall catalogue are flagged as blended. This is not a surprising result, as we expect resolved sources to have a naturally have higher polarisation fractions; they are located further from the centres of their host galaxies and are therefore less subject to in-situ depolarisation. Additionally, we have more resolving elements across an extended source, beam depolarisation is less an of an effect as the source size increases relative the beam.

\begin{table}
    \centering
    \begin{tabular}{lccc}
    \toprule
    Subset & Total count & Not blended count & Blended count\\
    \midrule
    All    & 3\,980\,618 & 3\,799\,614 & 181\,004\\
    \texttt{goodI} & 2\,350\,312 & 2\,093\,044 & 257\,268\\
    \texttt{goodRM} & 134\,401 & 54\,572 & 79\,829\\
    \bottomrule
    \end{tabular}
    \caption{
        Similar to Table~\ref{tab:counts}, except here the number counts are of PyBDSF sources in our catalogue having applied a subset or the\texttt{is\_blended\_flag}. Here the counts are derived from the exact same subsets as Table~\ref{tab:counts}. We count the number of sources by grouping the catalogue subsets by the \texttt{source\_id} column. We note that deduplication makes no difference to the source number counts.
    }
    \label{tab:source_counts}
\end{table}

\section{Analysis}\label{sec:analysis}

\subsection{Sensitivity and resolution}
\begin{figure*}
    \centering
    \begin{subfigure}{0.49\textwidth}
        \includegraphics[width=\textwidth]{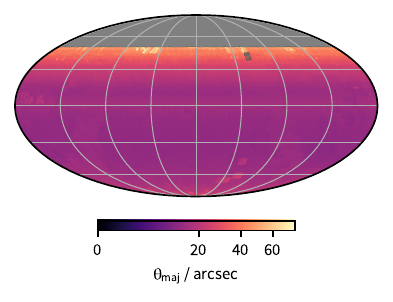}
        \caption{}\label{subfig:beam_maj}
    \end{subfigure}
    \begin{subfigure}{0.49\textwidth}
        \includegraphics[width=\textwidth]{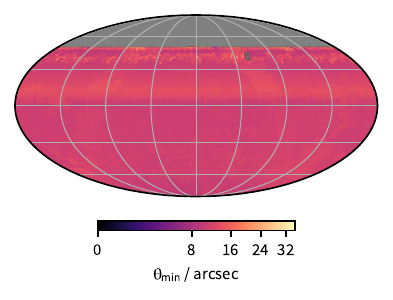}
        \caption{}\label{subfig:beam_min}
    \end{subfigure}
    \begin{subfigure}{0.49\textwidth}
        \includegraphics[width=\textwidth]{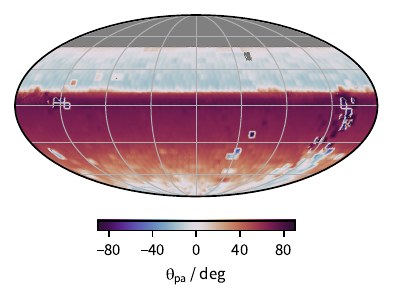}
        \caption{}\label{subfig:beam_pa}
    \end{subfigure}
    \begin{subfigure}{0.49\textwidth}
        \includegraphics[width=\textwidth]{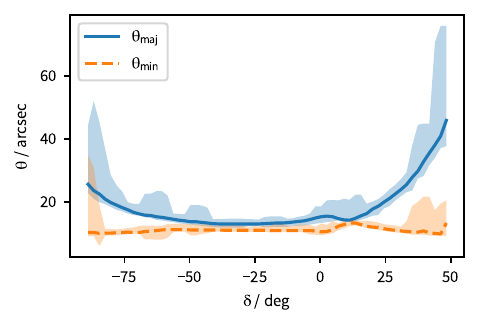}
        \caption{}\label{subfig:beam_dec}
    \end{subfigure}
    \caption{
        The point-spread function (PSF, $\theta$) across the survey area in celestial coordinates. We note that this PSF is the lowest common resolution across all channels in a given cutout cubelet. Panel (\subref{subfig:beam_maj}) shows the major axis ($\theta_\text{maj}$) with a square-root colour scale, panel (\subref{subfig:beam_min}) shows the minor axis also on a square-root colour scale, and panel (\subref{subfig:beam_pa}) shows the position angle on a linear scale. In panel (\subref{subfig:beam_dec}) we show the PSF specifically against the survey declination ($\delta$). The solid and dashed lines show the median $\theta_\text{maj}$ and $\theta_\text{min}$, respectively. We show the extrema of the these values with the shaded regions.
    }
    \label{fig:beam}
\end{figure*}

First, we begin by inspecting the spatial resolution across our survey area. In Figure~\ref{fig:beam} we show the survey PSF across the sky, and as a function of declination ($\delta$). This PSF represents the lowest common resolution across all channels and beams in the cutout around a given source. As discussed in \citetalias{Duchesne2023}, scheduling RACS observations within $\pm\qty{1}{\hour}$ of the meridian provides a spatially contiguous PSF that is primarily declination-dependent. Further, as a snapshot survey, the minor axis ($\theta_\text{min}$) and position angle ($\theta_\text{pa}$) both depend strongly on the instantaneous $uv$-coverage of the array for a given azimuth and elevation. For example, field-scale discontinuities seen in Figure~\ref{subfig:beam_pa} highlight where gaps in sky coverage, caused by avoiding solar system objects, were filled later in the survey. The major axis ($\theta_\text{maj}$) is also larger on the equator where Earth rotation synthesis no longer increases the $uv$-coverage. We also note that the array layout of ASKAP appears to create a circular PSF at $\delta\sim\ang{12}$.

In \citetalias{Hale2021} and \citetalias{Thomson2023} we elected to use a common resolution of \ang{;;25}, at the cost of sky area. By way of comparison, here our median PSF major axis is less than \ang{;;25} in the declination range $\ang{-87}<\delta<+\ang{30}$. Again we note, however, that we allow our angular resolution to vary for each source cutout. There are a small number of regions where flagging was relatively higher, producing poorer $uv$-coverage and angular resolution. Our poorest angular resolution occurs at the Northern declination limit of the survey. The median major axis in this region is $\theta_\text{maj}=\ang{;;45.9}$, with a maximum value of $\theta_\text{maj}=\ang{;;75.9}$. Across the majority of our survey area the angular resolution is much better, however, with a median and IQR of our major axis being $14.5 \substack{+5.5\\-1.4}\unit{arcsec}$.

\begin{figure*}
    \centering
    \begin{subfigure}{0.49\textwidth}
        \includegraphics[width=\textwidth]{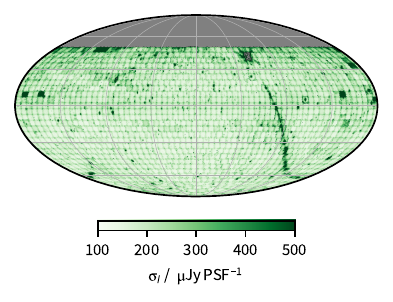}
        \caption{}\label{subfig:noise_I}
    \end{subfigure}
    \begin{subfigure}{0.49\textwidth}
        \includegraphics[width=\textwidth]{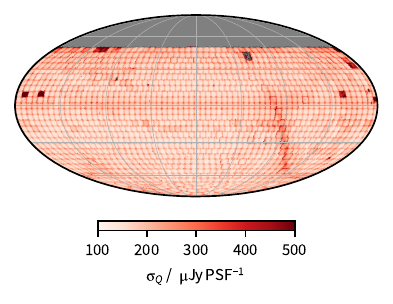}
        \caption{}\label{subfig:noise_Q}
    \end{subfigure}
    \begin{subfigure}{0.49\textwidth}
        \includegraphics[width=\textwidth]{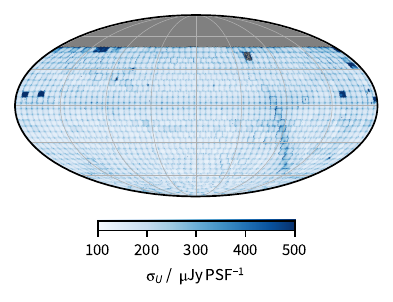}
        \caption{}\label{subfig:noise_U}
    \end{subfigure}
    \begin{subfigure}{0.49\textwidth}
        \includegraphics[width=\textwidth]{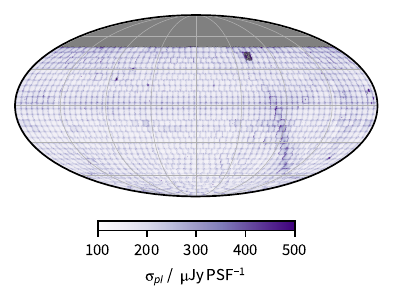}
        \caption{}\label{subfig:noise_P}
    \end{subfigure}
    \caption{
        Fitted, band-averaged $rms$ noise ($\sigma$) around each component across the survey area in celestial coordinates in (\subref{subfig:noise_I}) Stokes $I$, (\subref{subfig:noise_Q}) Stokes $Q$, (\subref{subfig:noise_U}) Stokes $U$, and (\subref{subfig:noise_P}) polarised intensity ($pI$). We note that value of $\sigma_{pI}$ is evaluated after performing RM-synthesis with inverse-variance weighting.
    }
    \label{fig:noise}
\end{figure*}

We now turn our attention to the survey sensitivity. As described in \S\ref{sec:cutout}, we extract the robust statistics for $rms$ noise ($\sigma$) of the pixels in the local vicinity of each extracted component. In Figure~\ref{fig:noise} we show the band-averaged $\sigma$ for Stokes $I$, $Q$, and $U$ as well as $pI$. We measure median and IQR noise across the sky of $211 \substack{+150\\-48}\,\si{\micro\jansky\per PSF}$, $194 \substack{+130\\-38}\,\si{\micro\jansky\per PSF}$, and $194 \substack{+140\\-40}\,\si{\micro\jansky\per PSF}$ in Stokes $I$, Stokes $Q$ and $U$, and $pI$, respectively. As expected, our average noise is marginally higher than the initial cataloguing due to both our more uniform weighting and the fact that we have not mosaicked adjacent fields. 

Inspecting the spatial distribution, we note three areas of increased noise. First, a handful of individual fields exhibit outlying noise across the entire observation. As noted in \S\ref{sec:data}, a number of target and calibration observations in the initial run of RACS-low3 were impacted by high amounts of flagging. This can be caused by either increased levels of RFI or by a dish rotation unwrap during an observation. As we have not made use of the re-observations, a number of fields in our set have poorer than expected sensitivity. We will be be able to rectify this in a future data release that uses all RACS-low3 observations. 

The second area of increased noise we note is in the vicinity of bright sources, particularly in Stokes $I$. As discussed in \citetalias{Duchesne2023}, the impact of bright sources can be significantly improved through peeling or visibility subtraction. Whilst we have not applied these techniques to this data release, we intend to do so in a future release along with other imaging and calibration improvements~(Galvin et al. in prep). 

The last area of increased noise in all Stokes parameters is along the Galactic plane. As a snapshot survey, RACS observations cannot recover extended emission with high fidelity, so we expect artefacts along the Galactic plane where bright, extended emission is common. This same extended emission will, to some degree, also be captured in our noise statistics where our noise annulus is much smaller than the scale of the emission itself. Finally, we remark on the lack of such outstanding noise outliers in the $pI$ noise distribution, which appears mostly uniform over the survey area. We can attribute this to our use of inverse-variance weighting in RM-synthesis.

\begin{figure}
    \centering
    \begin{subfigure}{\textwidth}
        \includegraphics[width=\textwidth]{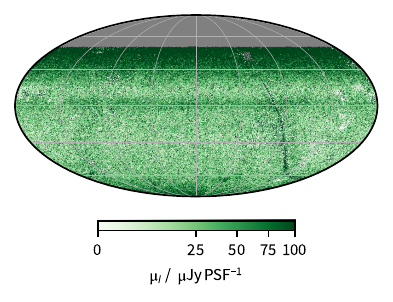}
        \caption{}\label{subfig:bkg_I}
    \end{subfigure}
    \begin{subfigure}{\textwidth}
        \includegraphics[width=\textwidth]{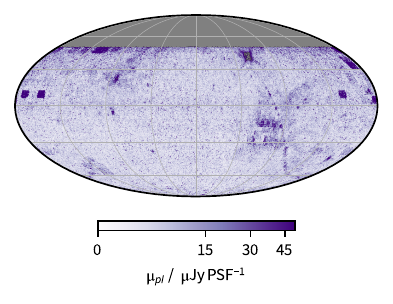}
        \caption{}\label{subfig:bkg_P}
    \end{subfigure}
    \caption{
        Fitted, band-averaged background ($\mu$) around each component across the survey area in celestial coordinates in (\subref{subfig:bkg_I}) Stokes $I$ and (\subref{subfig:bkg_P}) polarised intensity ($pI$). We note that both sub-figures use a square-root colour scale.
    }
    \label{fig:bkg}
\end{figure}

In addition to the $rms$ noise around each component, we also simultaneously compute the background ($\mu$) and subtract this channel-wise from each Stokes spectrum (see \S\ref{sec:cutout}). On average across the sky and band we find backgrounds of $17 \substack{+110\\-35}\,\si{\micro\jansky\per PSF}$, $0 \substack{+19\\-18}\,\si{\micro\jansky\per PSF}$, and $0 \substack{+18\\-18}\,\si{\micro\jansky\per PSF}$ in Stokes $I$, $Q$, and $U$, respectively. In Figure~\ref{fig:bkg} we show how the band-averaged background is distributed across the sky in Stokes $I$ and $pI$. Inspecting Stokes $I$ we see that the background is predominantly declination-dependent, with some enhancement along the Galactic plane. Since the spatial distribution of this statistic is clearly correlated with the survey $uv$-coverage and resulting PSF, we posit that this statistic is mostly capturing residual sidelobe noise in the Stokes $I$ images. In the $pI$ background we can again see the impact on the handful of fields impacted by RFI. Remarkably, we can see diffuse polarised Galactic features, particularly the North Polar Spur, which are commonly seen in single-dish surveys. \citet{Rudnick2009} previously recovered such features from a re-processing of NVSS. We leave detailed analysis of these structures to future work, however we can draw some basic conclusions on the basis of our $uv$ coverage. As noted in \S\ref{sec:imaging}, we apply a global $uv$-cut of $200\,\lambda$ which corresponds to largest angular scale of $\sim\ang{;17}$. These Galactic polarised emission features must therefore be composed of structures with spatial scales less than this angular size.

\subsection{Flux density accuracy}
To validate our flux density accuracy in total intensity and linear polarisation we obtain calibrator measurements with the Very Large Array (VLA) from \citet{Perley2013} and \citet{Perley2017} (hereafter \citetalias{Perley2013} and \citetalias{Perley2017}), as well as with MeerKAT from \citet[][hereafter \citetalias{Taylor2024}]{Taylor2024}. These data are presented as integrated flux densities, whereas we only extract pixel-wise peak flux densities. These values should, ideally, be the same for unresolved sources; however, the same will not be true for extended sources. Further, there may be some small differences between our pixel-wise peak and a fitted peak, with the pixel-wise peak being systematically lower. We must also be aware that fractional polarisations can change over time~\citepalias{Perley2013}, and can be strongly frequency-dependent. Given these constraints, and the intended purpose of this work as a rotation measure grid, we leave detailed analysis of our flux scale to future work (Galvin et al. in prep.) and we will simply validate our flux scale.

For reference, \citetalias{Perley2017} provide an integrated flux density scale for 20 sources in the range of \qty{50}{\mega\hertz} to \qty{50}{\giga\hertz}, as well as the largest angular scale (LAS) of each source. In polarisation \citetalias{Perley2013} report on four sources (3C48, 3C138, 3C147, and 3C286) in the frequency range of \qtyrange{1}{50}{\giga\hertz} at the epoch of December 2010. The work of \citetalias{Taylor2024} provide a curated list of 10 selected broad-band polarisation calibrator sources (out of an initial 98 sources). They present their total intensity and polarisation information at a reference frequency of \qty{1.4}{\giga\hertz} with scaling parameters and standard errors.

\begin{figure*}
    \includegraphics{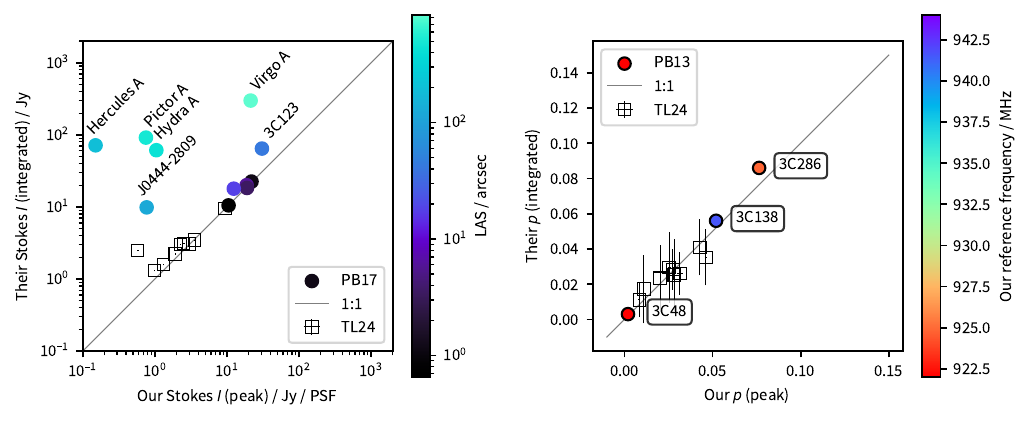}
    \caption{
        Comparison of Stokes $I$ flux density (left) and polarisation fraction (right) against calibration sources. In both cases, our values are derived from the peak pixel on the source whereas the reference values are integrated. In total intensity, we compare our peak flux density against fitted models from both \citetalias{Perley2017} (circles) and \citetalias{Taylor2024} (squares) evaluated at our reference frequency. The colour scale represents the largest angular size (LAS) as reported by \citetalias{Perley2017}. Where the $\text{LAS}>\ang{;;20}$ we label the source name. In fractional polarisation we compare against the measurements from \citetalias{Perley2013} (circles) and \citetalias{Taylor2024} (squares). We use the fractional values from \citetalias{Perley2013} at \qty{1.050}{\giga\hertz}, whereas we use the depolarisation term from \citetalias{Taylor2024} to scale to our reference frequency. We colour the points from \citetalias{Perley2013} by our reference frequency and label them with their source name. We show all error bars at $5\sigma$.
    }
    \label{fig:flux}
\end{figure*}

We begin our comparison by crossmatching (symmetrically) the calibrator positions to our de-duplicated catalogue. Taking a maximum angular separation of \ang{;;20}, we find 11 matches with \citetalias{Perley2017}, 9 matches with \citetalias{Taylor2024}, and 3 sources from \citetalias{Perley2013}. In total intensity we evaluate the models from both \citetalias{Perley2017} and \citetalias{Taylor2024} at our own reference frequency. We show the comparison of the total intensity flux density in the left panel of Figure~\ref{fig:flux}. Excluding sources from \citetalias{Perley2017} with an $\text{LAS} > \ang{;;20}$, we find that our total intensity is in agreement with the calibrator measurements within $11\%$. In polarisation fraction we select the \qty{1.050}{\giga\hertz} data from \citetalias{Perley2013} (which lies within our observed band), and scale the \citetalias{Taylor2024} fraction to our reference frequency with their reported depolarisation term. We show the comparison of the matched fractional polarisations in the right-hand panel of Figure~\ref{fig:flux}. The median fractional difference between our measurements and the calibrators is \qty{16}{\percent} with a MADstd of \qty{0.4}{\percent}. Given the above caveats, we are satisfied that our flux scale is accurate to $\sim\qty{15}{\percent}$; which is in line with the previous epochs of RACS~\citepalias{Hale2021,Duchesne2023,Duchesne2025}.

As a final comparison of our polarisation calibration, we also compare our measured polarisation angles ($\chi$) against the \citetalias{Perley2013} and \citetalias{Taylor2024} calibrator surveys. Due to the very Faraday rotation effect we are interested in, the polarisation angle is strongly frequency dependent. As such we de-rotate all polarisation angles to a common reference frequency of \qty{1}{\giga\hertz}, which is in-band for all three surveys. We de-rotate assuming a single RM:
\begin{equation}
    \chi = \chi_r + \text{RM} \left( \lambda^2 - \lambda_r^2 \right),
\end{equation}
where $\chi_r$ and $\lambda_r^2$ are the polarisation angle and wavelength-squared at the reference frequency. As such, any Faraday complexity will result in additional scatter of the de-rotated angle. 

\begin{figure}
    \centering
    \includegraphics[width=\textwidth]{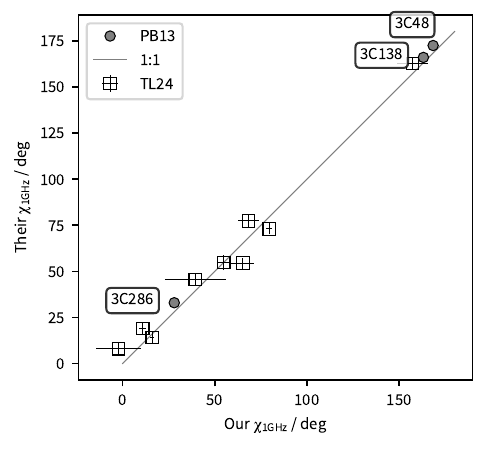}
    \caption{
        Comparison of polarisation angles ($\chi$) against calibration sources (as per Figure~\ref{fig:flux}). Data from \citetalias{Perley2013} and \citetalias{Taylor2024} are shown in circles and squares, respectively. Here we have re-rotated all angles to a common reference frequency of \qty{1}{\giga\hertz} using the reported rotation measures from each catalogue. We show error bars at $5\sigma$.
    }
    \label{fig:angle_check}
\end{figure}

We show the comparison of our polarisation angles against the calibrator survey values in Figure~\ref{fig:angle_check}. Against just the data separately from \citetalias{Perley2013} and \citetalias{Taylor2024} the mean and standard deviation of the difference in measured angles is $\ang{4.0}\pm\ang{0.9}$ and $\ang{2}\pm\ang{7}$, respectively. Combined the difference in angles is $\ang{3}\pm\ang{6}$. We conclude overall that our absolute polarisation angle calibration is accurate within $\sim\ang{5}$.

\begin{figure*}
    \includegraphics{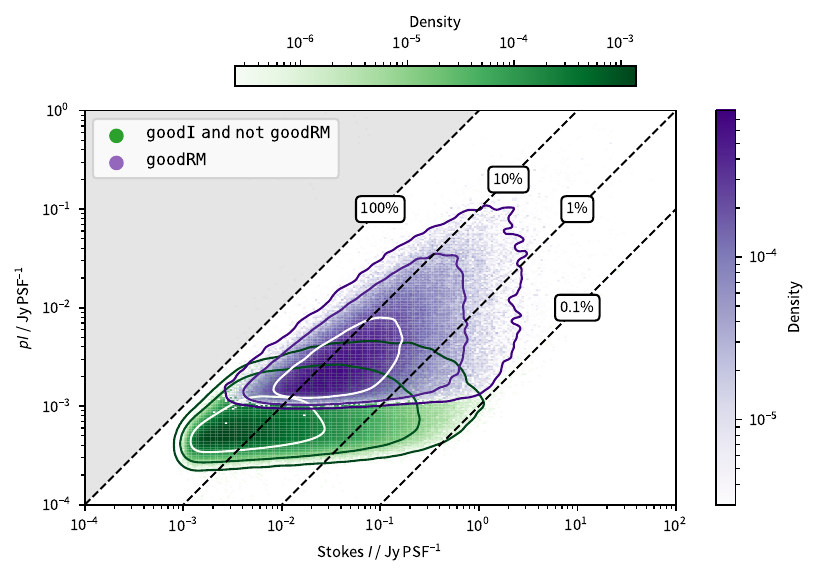}
    \caption{
        Two-dimensional histogram of the Stokes $I$ distribution against $pI$ from our concatenated catalogue. In green we show the density of components from our \texttt{goodI} and not \texttt{goodRM} subset (see \S\ref{sec:subsets}), and in purple we show the subset where \texttt{goodI} and \texttt{goodRM} are true. We show contours at the $16^\text{th}$, $50^\text{th}$, and $84^\text{th}$ percentiles. In dashed lines we show regions of constant fractional polarisation. We shade the forbidden region of over $100\%$ fractional linear polarisation in grey.
    }
    \label{fig:total_polint}
\end{figure*}

Having satisfied ourselves with the flux scale, we can inspect the distribution of our total and linearly polarised flux density. In Figure~\ref{fig:total_polint} we show the 2D histogram of Stokes $I$ against $pI$ for both our \texttt{goodI} and \texttt{goodRM} subsets. For the following values we will report the median with upper and lower ranges given by the $16^{\text{th}}$ and $84^{\text{th}}$ percentiles. In the overall \texttt{goodI} subset we find a distribution of $I=6.0 \substack{+20\\-3.6}\,\unit{\milli\jansky\per PSF}$, and $I/\sigma_I=26 \substack{+82\\-15}$. Where \texttt{goodI} applies but not \texttt{goodRM} the sample is overall lower in Stokes $I$ with $I=5.4\substack{+13\\-3.0}\,\unit{\milli\jansky\per PSF}$ and $I/\sigma_I=24 \substack{+56\\-12}$. We can also see the $pI$ is systemically lower for this subset, as expected from our polarised signal-to-noise and leakage cuts. In the $\texttt{goodRM}$ subset, the distribution in total intensity is $I=55 \substack{+130\\-35}\,\unit{\milli\jansky\per PSF}$ and $I/\sigma_I=230 \substack{+490\\-150}$. In polarised intensity we find $pI=2.9 \substack{+3.9\\-1.2}\,\unit{\milli\jansky\per PSF}$ with $pI/\sigma_{pI}=14.0 \substack{+18\\-4.8}$. In fractional polarisation the distribution is $p=6.1 \substack{+6.4\\-3.7}$.

\begin{figure}
    \centering
    \begin{subfigure}{\textwidth}
            \includegraphics[width=\textwidth]{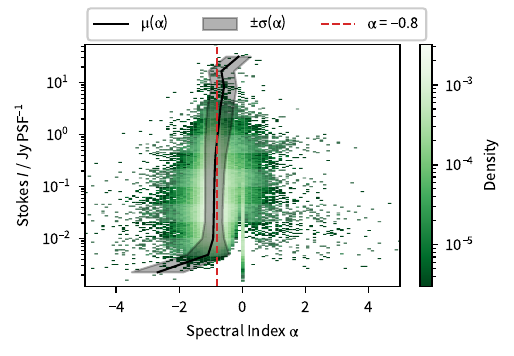}
            \caption{}\label{subfig:index_vs_flux}
    \end{subfigure}
    \begin{subfigure}{\textwidth}
            \includegraphics[width=\textwidth]{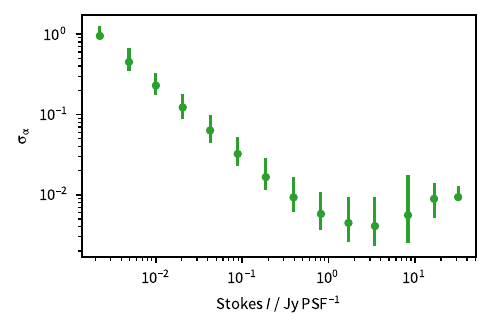}
            \caption{}\label{subfig:index_err_vs_flux}
    \end{subfigure}

    \caption{
        Stokes $I$ spectral indices ($\alpha$). In (\subref{subfig:index_vs_flux}) we show the 2D histogram of $\alpha$ against Stokes $I$ flux density from our concatenated catalogue in the range $-5\leq\alpha\leq5$. In the black solid and shaded region we show the error-weighted mean ($\mu$) and standard deviation ($\sigma$) of the $\alpha$ in bins of flux density. In the vertical red dashed line we show where $\alpha=-0.8$. We note that, due to our hierarchical fitting method, some models are fit with a flat spectral index, which effectively sets $\alpha=0$. This results in the sharp feature in the distribution at $\alpha=0$. In (\subref{subfig:index_err_vs_flux}) we show the error on the spectral index $\sigma_\alpha$ as function of Stokes $I$. Each error bar shows the median and with 16$^\text{th}$ and 84$^\text{th}$ percentile range.
    }
    \label{fig:spectral_index}
\end{figure}

Finally, we can inspect our fitted in-band spectral indices ($\alpha$). We find an error-weighted median of $\alpha=-0.8\pm0.4$; consistent with both previous surveys of extragalactic sources~\citep[e.g.][]{Condon1992} and the multi-band analysis of RACS from \citetalias{Duchesne2025}. In Figure~\ref{subfig:index_vs_flux} we show the distribution of $\alpha$ against total flux density. Overall, there is not a strong error-weighted trend in $\alpha$ with flux density until $I\gtrsim\qty{10}{\jansky\per PSF}$ where $\alpha$ begins to flatten on average. We should also note that a \qty{288}{\mega\hertz} band at $\sim\qty{1}{\giga\hertz}$ is not a particularly long lever arm for determining $\alpha$. As we show in Figure~\ref{subfig:index_err_vs_flux}, below $I\lesssim\qty{100}{\milli\jansky\per PSF}$ the average error on $\alpha$ rises quickly to $\sigma_\alpha\sim1$. Above $I\gtrsim\qty{100}{\milli\jansky\per PSF}$ the $\sigma_\alpha$ is on the order of $10^{-2}$. However, as described above, half of our components in the \texttt{goodI} subset have a flux density $I<\qty{6}{\milli\jansky\per PSF}$. Overall, our spectral indices are physically reasonable and certainly sufficient for fractional polarisation analysis through RM synthesis. Users of our reported spectral indices will need to pay careful attention to our reported errors when using them for scientific analysis.

\subsection{Residual polarisation leakage}\label{sec:leakage}
As with all widefield ASKAP images, residual widefield polarisation leakage is a critical systemic that can limit the resulting density of our RM grid~\citepalias{Thomson2023}. Further, we also require a method of measuring our residual leakage after correction so that we can successfully select sources with astrophysical, rather than instrumental, linear polarisation. As described in \S\ref{sec:data}, our set of RACS-low3 fields were observed entirely with a single set of beamforming weights. This set of weights had holographic measurements made of the beam that we used to correct for the primary beam and off-axis leakage from Stokes $I$ into $Q$ and $U$.

In a similar process to \citetalias{Thomson2023}, we are able to measure the residual leakage across the field of view of observations by using field sources. In using this technique, we are making the assumption that a majority of sources are not intrinsically linearly polarised. Unlike in DR1, since our data used a single set of beamforming weights, we have the ability to stack across our entire dataset for this measurement.

\begin{figure*}
    \centering
    \begin{subfigure}{0.49\textwidth}
        \includegraphics[width=\textwidth]{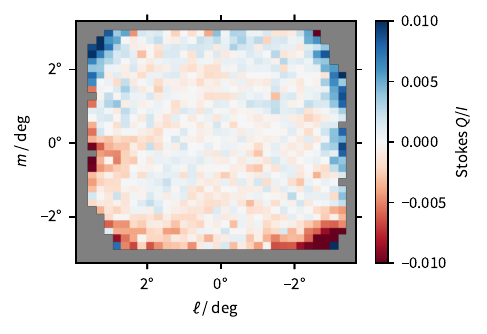}
        \caption{}\label{subfig:q_leak}
    \end{subfigure}
    \begin{subfigure}{0.49\textwidth}
        \includegraphics[width=\textwidth]{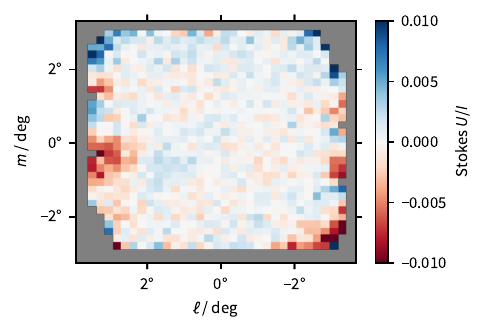}
        \caption{}\label{subfig:u_leak}
    \end{subfigure}
    \begin{subfigure}{0.49\textwidth}
        \includegraphics[width=\textwidth]{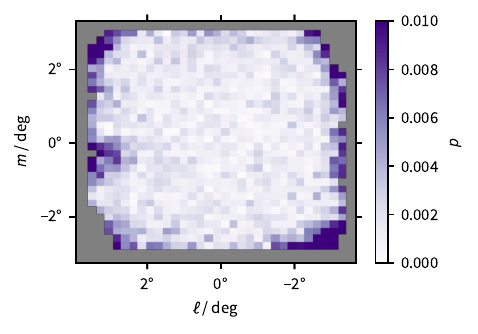}
        \caption{}\label{subfig:p_leak}
    \end{subfigure}
    \begin{subfigure}{0.49\textwidth}
        \includegraphics[width=\textwidth]{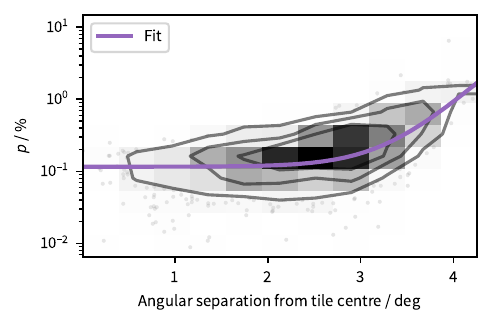}
        \caption{}\label{subfig:sep_leak}
    \end{subfigure}
    \caption{
        Residual leakage across the field of view in RACS-low3. Panels (\subref{subfig:q_leak}) and (\subref{subfig:u_leak}) show our estimate of the residual leakage from Stokes $I$ into Stokes $Q$ and $U$ in the telescope frame, respectively. We combine these to produce the leakage in fractional polarisation ($p$), which we show in panel (\subref{subfig:p_leak}). We flatten this fractional map to view the residual leakage as a function of angular separation from the tile centre, which we show in panel (\subref{subfig:sep_leak}). In this panel we give the fractional leakage as a percentage on a logarithmic scale, and we produce a 2D histogram where the points are over-dense. In the solid purple line we show our empirical fit to the residual leakage which has the functional form: $ae^{b\vartheta}+c$ where $\vartheta$ is the angular separation. Our fitted values are $a=2.7\times10^{-7}$, $b=2.58$, and $c=1.16\times10^{-3}$.
    }
    \label{fig:leakage}
\end{figure*}

We begin stacking by selecting high signal-to-noise Stokes $I$ sources from our full catalogue where $I/\sigma_I > 20$. We then form a grid across the instrument field of view in $l,m$ taking 31 bins in each dimension, resulting in bins of $\ang{;13.5}\times\ang{;11.5}$ in size. To minimise the impact of astrophysically polarised sources we first filter using a sigma-clip, and then take the median of the resulting clipped values. We use \verb|astropy.stats.sigma_clip| to perform the sigma-clip down to a level of $3\sigma$ with no limit on the number of iterations, and we use the median absolute deviation (MAD) normalised to the standard deviation (MADstd, using \verb|astropy.stats.mad_std|). We show the resulting distributions of Stokes $Q/I$ and $U/I$ in Figures~\ref{subfig:q_leak} and \ref{subfig:u_leak}, respectively. Inspecting these Figures we can see what appears to be an apparent shift along the $m$ axis, resulting in higher residual leakage towards the edges of the field of view with anti-symmetry about $m=\ang{0}$. We found a similar effect in \citetalias{Duchesne2023} in the apparent brightness of total intensity sources. There, we also found a flip in the apparent shift depending on where in the sky a given set of primary beams where determined and then applied. Here, however, we find that further sub-dividing our catalogue reduces the residual leakage `signal' to a point where concrete analysis is no longer viable. Despite this, it is still worthy of note that our data again shows this affect with ASKAP data, and should be noted for future surveys and observations.

To determine the leakage from $I$ into $p$ we form the polarised intensity surface from our $Q/I$ and $U/I$ grids, which we show in Figure~\ref{subfig:p_leak}. For the purposes of our catalogue we collapse the $l,m$ axes into a simple angular separation ($\vartheta$) and bin the resulting values to minimise the impact of variance in our leakage surface. Finally, we fit an empirical model to this data of the functional form:
\begin{equation}
    p_\text{leak} = a e^{b\vartheta} +c,
\end{equation}
where our fitted values are $a=2.7\times10^{-7}$, $b=2.58$, and $c=1.16\times10^{-3}$. This informs us that, over most of the field of view, the residual leakage is $\sim\qty{0.1}{\percent}$. This is an order of magnitude improvement over our DR1 result~\citepalias{Thomson2023}. Beyond a separation of $\sim\ang{2.5}$ from the tile centre the residual leakage climbs to $\sim\qty{1}{\percent}$. For each row in our catalogue we evaluate the fitted model and report the resulting value in the \texttt{leakage} column.

\subsection{Rotation measure accuracy and analysis}\label{sec:rm_accuracy}
To validate our determined RMs we compare with a collection of historical RM surveys. We begin with version 1.2.0 of the \textsc{RMTable} compilation~\citep[][hereafter \citetalias{VanEck2023}]{VanEck2023}. To this we add version 9 of the ATNF Pulsar Catalogue SQLite database\footnote{\url{https://doi.org/10.25919/jr11-yj30}}~\citep[][hereafter \citetalias{Manchester2005}]{Manchester2005,Hobbs2025}, as well as catalogues from \citetalias{Thomson2023}, \citetalias{Taylor2024}, and \citet{Loi2025}. We cross-match this combined collection symmetrically with our de-duplicated, \texttt{goodRM} catalogue using \pkg{astropy.coordinates.SkyCoord.match\_to\_catalog\_sky} with a maximum separation based on the angular resolution of the respective surveys. We take the maximum of either our own or the external catalogue's PSF major axis, and set the maximum angular separation to be twice that value.

Despite their common format (thanks to the \textsc{RMTable} schema), the construction of each catalogue may be vastly different. For example, the nominal frequency coverage, the RM algorithm used, and application of ionospheric correction may all differ greatly between catalogues. Further, true RM variation may occur between the epochs of each component observation~\citep[e.g.][]{Anderson2019}. As such we perform our comparison on the basis of each individual catalogue in our collection. For the purpose of comparison we will be analysing the distribution of differences between our RM and the historical RM ($\Delta\text{RM}$) and the error on this difference ($\sigma_{\Delta\text{RM}}$). If the variance in $\Delta\text{RM}$ was purely caused by Gaussian noise, and if the errors are correctly measured, we would expect to see an error-normalised difference in RM of $\Delta\text{RM}/\sigma_{\Delta\text{RM}}=0\pm1$. The error normalised difference between any pair of uncorrelated values ($X$) is given by
\begin{equation}
    \frac{\Delta X}{\sigma_{\Delta X}} = \frac{X_1 - X_2}{\sqrt{\sigma_{X_1}^2 + \sigma_{X_2}^2}},
    \label{eqn:error_diff}
\end{equation}
where here we substitute $X=\text{RM}$ and subscripts $1$ and $2$ denote the first and second components in a pair, respectively.

By far the largest catalogue in our historical collection, by both number of RMs and areal coverage, is from \citetalias{Taylor2009}. Since RMs in this catalogue were computed from two narrowly-spaced frequencies, the reported RMs may suffer from the $n\pi$ ambiguity. Indeed, this has been found to be the case; \citet{Ma2019} report a lower limit of 50 RMs. A single-wrap ($n=1$) error corresponds to $\pm\qty{652.9}{\radian\per\metre\squared}$. Upon cross-matching with our catalogue, in addition to components which are centred $\Delta\text{RM}=\qty{0}{\radian\per\metre\squared}$, we find two populations centred on expected $1\pi$ wrap error. As such, we make a correction to the matched \citetalias{Taylor2009} catalogue; we take any component where $|\Delta\text{RM}|>\qty{400}{\radian\per\metre\squared}$ and subtract $\pm\qty{652.9}{\radian\per\metre\squared}$ (with the appropriate sign) to bring all \citetalias{Taylor2009} components into a single distribution around $\Delta\text{RM}=\qty{0}{\radian\per\metre\squared}$. Overall, we find $30\,183$ matching components with \citetalias{Taylor2009}, 146 of which we correct for the $n\pi$ ambiguity.

\begin{figure}
    \centering
    \includegraphics[width=\columnwidth]{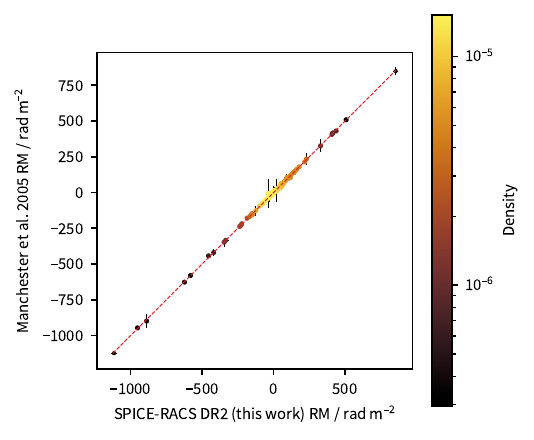}
    \caption{
        Our rotation measures (RM) cross-matched to 127 pulsar RMs from \citetalias{Manchester2005}. Where scatter points are over-dense we show the density of points in the colourscale. We show the 1:1 relation in the red, dashed lined.
    }
    \label{fig:pulsar}
\end{figure}

Having queried the \citetalias{Manchester2005} database, we retrieve 3595 pulsars with reported coordinates. After cross-matching with our catalogue, and removing duplicate matches for multiple pulsars which match the same RACS component, we find 133 matches to known pulsars. Of these, 6 do not have reported RMs; we show our RMs, along with our component name and the pulsar name, in Table~\ref{tab:pulsar}. We show the distribution of pulsar RMs against our own for the remaining 127 pulsars in Figure~\ref{fig:pulsar}. Initially, we found that a single pulsar, J1406-5806, was outlying from the 1:1 line. On closer inspection, we find that the offset from our measured RM is almost exactly the twice magnitude of the RM. As such, we assume this to be a sign error from the original catalogue of \citet{Kramer2003} and we invert the sign of the pulsar value to match our own. Having made this single correction, the overall distribution of RM between the two datasets is very tight, with the error-normalised difference in RM $\Delta\text{RM}/\sigma_{\Delta\text{RM}}=0.4 \substack{+1.7\\-1.5}$.

\input{tables/new_pulsars}

\begin{figure*}
    \centering
    \includegraphics[width=\columnwidth]{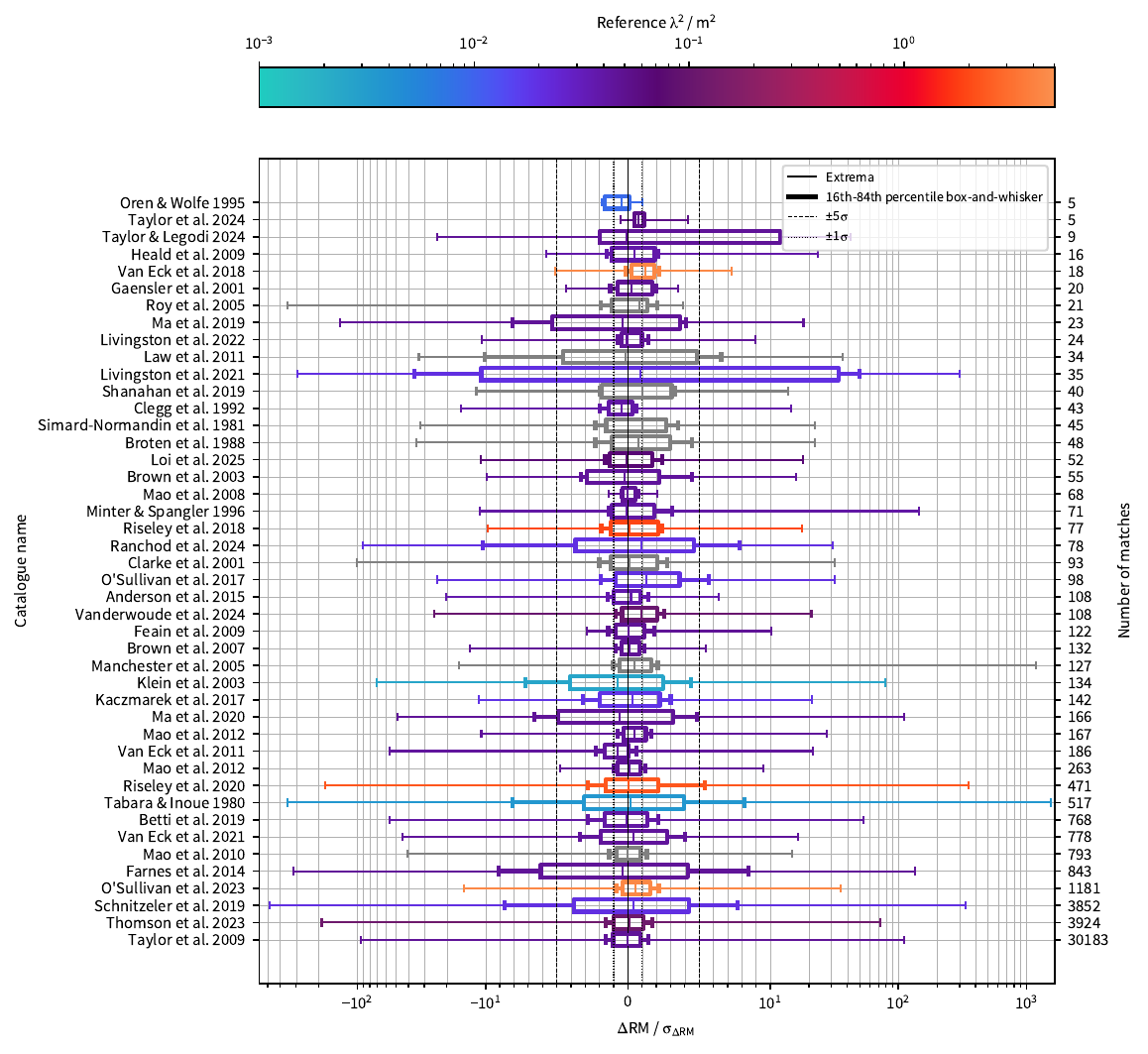}
    \caption{
        The difference in RM ($\Delta\text{RM}$) between SPICE-RACS DR2 matched with an external catalogue, normalised by the error in the $\Delta\text{RM}$ ($\sigma_{\Delta\text{RM}}$). The majority of catalogues listed above are components of \citetalias{VanEck2023} v1.2.0. To this we have added the catalogues from \citetalias{Thomson2023}, \citetalias{Taylor2024}, \citet{Loi2025}, and the \citetalias{Manchester2005}~(\citet{Manchester2005} in the axis label). We present the distribution of the $\Delta\text{RM}/\sigma_{\Delta\text{RM}}$ for each catalogue as a box-and-whisker diagram~\citep{Spear1952, Tukey1977}. In the thick lines we show the box-and-whisker distribution between the 16th and 84th percentiles. In the thin lines we show the extrema of the distribution. The vertical dotted and dashed lines show the $\pm1\sigma$ and $\pm5\sigma$ ranges, respectively. For perfectly normally-distributed errors we would expect the inner box of each distribution to lie within the $\pm1\sigma$ range. We have sorted each catalogue by the number of matches with SPICE-RACS DR2 from most at the bottom to least at the top, with the number of matches listed on the right-hand axis. The corresponding authorship name for each catalogue is given on the left axis. We also colour each distribution by its corresponding reference wavelength-squared ($\lambda^2$) with a logarithmic colour scale. We note that same catalogues do not contain a single, homogeneous reference wavelength-squared which we cannot represent in this figure. If a catalogue does have reported reference wavelength-squared, we colour the lines grey. Note that the references for the sub-catalogues of \citetalias{VanEck2023} are given in \citet{VanEck2023}. We also enumerate all such references in our own reference list.
    }
    \label{fig:crossmatch}
\end{figure*}

Having made the correction described above, we now look to comparing our RMs with the historical compilation in bulk. In Figure~\ref{fig:crossmatch} we show the distribution of $\Delta\text{RM}/\sigma_{\Delta\text{RM}}$ for each catalogue. In this Figure we also denote the number of crossmatches to our components, as well as the nominal reference frequency of each catalogue. Deep investigation of each matched component is beyond the scope of this work, however we note a number of interesting features. Overall, our measured RMs are in bulk agreement with the historical compilation, with a handful of outliers in each catalogue. 

To investigate whether these outliers are caused by systematics we first repeat our cross-matching analysis, taking our reference catalogue to be \citetalias{Taylor2009} and then \citetalias{Schnitzeler2019}. We find similar results as when we use SPICE-RACS DR2 as the reference catalogue. A deep analysis of the differences between each catalogue is beyond the scope of this work, but there are a few historical features to note. Catalogues which sample relatively few values of $\lambda^2$, such as \citetalias{Taylor2009}, as subject to $n\pi$ ambiguity errors. The nominal $\lambda^2$ coverage, which we indicate in Figure~\ref{fig:crossmatch}, can affect which RM components are detected. For example, at long wavelengths components can become depolarised, whereas at short wavelength multiple components can become unresolved. Additionally, some RM components detected with sufficient signal-to-noise can be observed to vary over time~\citep{Anderson2019}.

We note a larger number of outlying RMs in comparison with \citetalias{Schnitzeler2019}, the previous largest Southern sky RM survey. We can suggest a few reasons for disagreement between our RM values. First, the angular resolution of \citetalias{Schnitzeler2019} is relatively coarse ($\sim\ang{;2}$) which makes correctly matching components difficult. Second, the higher observed band (\qtyrange{1}{3}{\giga\hertz}) means that \citetalias{Schnitzeler2019} is less sensitive to depolarisation effects, and may be finding a different peak in the Faraday spectrum. Additionally, \citetalias{Schnitzeler2019} also used their own $QU$-fitting approach applied directly to visibilities, rather than RM-synthesis of extracted image spectra. Finally, inspecting the distribution of RMs of \citetalias{Schnitzeler2019} against both our own RMs and \citetalias{Taylor2009}, we see that \citetalias{Schnitzeler2019} reports many more RMs above $|\text{RM}|\gtrsim\qty{1000}{\radian\per\metre\squared}$. Such high RM values likely cause strong depolarisation in our sub-\qty{1}{\giga\hertz} spectra.

The catalogue with the largest bulk difference from our values is from \citet{Livingston2021}. This work targeted sources towards the Galactic centre, and found significant Faraday complexity in their sample. In effect, this large dispersion of $\Delta\text{RM}/\sigma_{\Delta\text{RM}}$ complements their in-band Faraday complexity result. These sources would make excellent targets for follow up monitoring observations and searches of astrophysical variability and Faraday complexity.

As final check for systematic issues in our own catalogue, we perform a statistical feature comparison to identify which columns in our matched catalogue are predictors for large differences in RM relative to matched catalogues. We split our catalogue in two groups based on the value $\Delta\text{RM}/\sigma_{\Delta\text{RM}}$.

For numerical values we obtain a $p$-value from a Mann-Whitney U test~\citep[][using \pkg{scipy.stats.mannwhitneyu}]{Mann1947} and compute the Cohen's $d$ value~\citep{Cohen1988} to identify significant columns. Similarly for `categorial' columns (such as the matched external catalogue name), we obtain a $p$-value from a chi-squared test (using \pkg{scipy.stats.chi2\_contingency}) and Cram\'er's $V$ value. We then correct the $p$-values for multiple test false detections with Benjamini-Hochberg~\citep{Benjamini1994} correction \citep[using \pkg{statsmodels.stats.multitest.multipletests} from][]{seabold2010statsmodels}. We then filter the resulting tests using $p<0.05$ combined with either $d\geq0.5$ or $V\geq0.1$. Here we take the $d$ and $V$ values to measure the effect size for numerical and categorial columns, respectively.

To we select a value of $\Delta\text{RM}/\sigma_{\Delta\text{RM}}=5$ on which to split our matched catalogue. Following the above filtering, we find the following columns are statistically significant between the two groups (sorted by effect size): \texttt{other\_reffreq\_pol}, \texttt{stokesI}, \texttt{sigma\_add}, \texttt{rm\_width}, \texttt{peak\_I\_flux\_err}, \texttt{peak\_I\_flux}, \texttt{total\_I\_flux\_err}, \texttt{total\_I\_flux}, \texttt{rm\_diff\_err}, \texttt{fdf\_noise\_mad}, \texttt{other\_rm\_err}, \texttt{snr\_polint}, \texttt{fracpol}, and \texttt{other\_cat\_name}. We see that the strongest indicator of effect size is the reference frequency of the catalogue we match to. Additionally, our RM-complexity metrics are two of the strongest predictors of difference between the two groups. We discuss our complexity metrics in detail in \S\ref{sec:complex}, however here we will note that we expect all the columns list here to correlate strongly with Faraday complexity except for \texttt{rm\_diff\_err} (which is $\sigma_{\Delta\text{RM}}$) and \texttt{other\_cat\_name} (the name of the external catalogue). 

The sign of the $d$-value also indicates the relative magnitude of parameters between our subsets. For our test, a positive $d$-value indicates the parameter is larger in in the $\Delta\text{RM}/\sigma_{\Delta\text{RM}}>5$ set, whereas a negative value indicates it is smaller. Of the catalogue columns listed above only \texttt{rm\_diff\_err}, \texttt{other\_rm\_err}, and \texttt{fracpol} have a negative $d$-value, whereas all others are positive. Smaller values of \texttt{rm\_diff\_err} and \texttt{other\_rm\_err} could indicate either underestimated errors in the external catalogue. Such a case could arise from low-frequency observations where the narrow RMSF provides highly precise RMs such that the error budget becomes dominated by systematic errors which are harder to quantify. Overall, however, the $d$-value of \texttt{other\_reffreq\_pol} is positive, indicating that the reference frequency of the external catalogue is larger than ours in the $\Delta\text{RM}/\sigma_{\Delta\text{RM}}>5$. As discussed above, higher frequency observations are more sensitive to Faraday thick features, meaning they will measure measure a different RM in the presence of depolarisation effects \citep[see discussion in e.g.][]{VanEck2017}. Finally, we note smaller values of \texttt{fracpol} in the $\Delta\text{RM}/\sigma_{\Delta\text{RM}}>5$ set. This difference, combined with larger relative values of total intensity, could be an indicator of residual polarisation leakage in our catalogued RMs. We investigate the internal reliability of our RM catalogue in \S\ref{sec:time}.

Overall, we conclude that the majority of the outliers we see are caused by systematic differences between our survey and historical catalogues, or multiple components in our Faraday spectrum, as indicated by Faraday complexity.

\subsubsection{SPICE-RACS DR2 RM grid}\label{sec:rm-grid}

\begin{figure*}
    \includegraphics{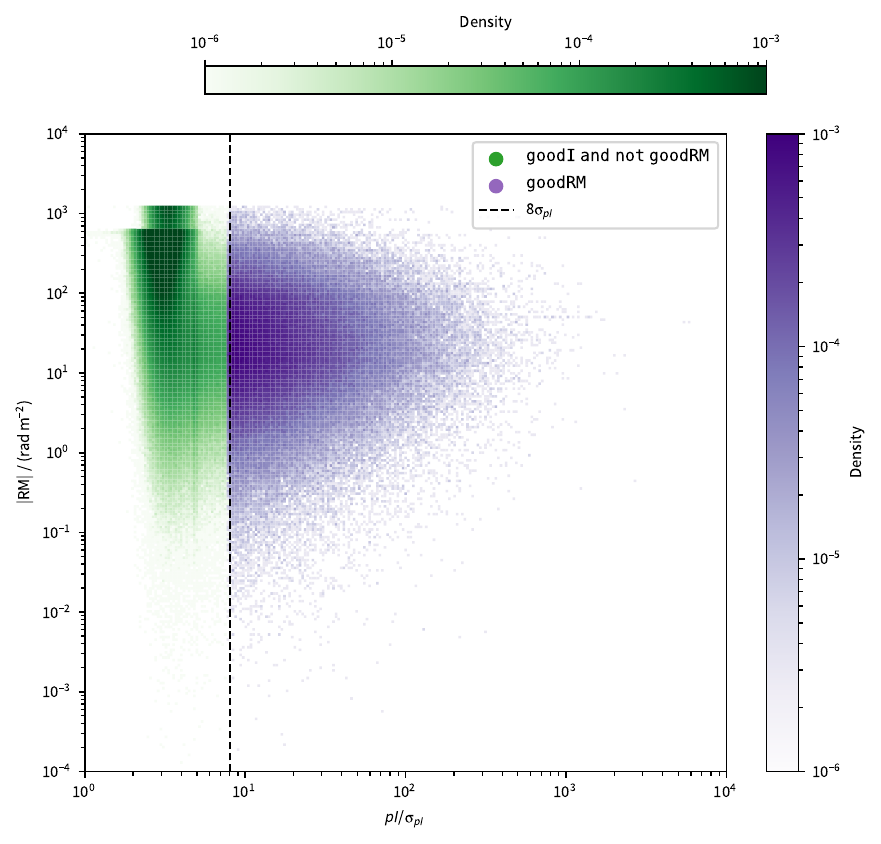}
    \caption{
        Two-dimesional histograms of the absolute value of rotation measure ($|\text{RM}|$) against polarised intensity signal-to-noise ($pI/\sigma_{pI}$). We colour the region where our \texttt{goodRM} subset applies in purple, and in green we shade where \texttt{goodI} applies but not \texttt{goodRM} (see \S\ref{sec:subsets}). In the vertical dashed line we show divider between \texttt{goodI} and \texttt{goodRM} where $pI/\sigma_{pI}=8$.
    }
    \label{fig:rm_polint}
\end{figure*}

Having validated our RMs against previous observations, we now look at our own dataset directly. First, we we look at the distribution of our RMs as a function of polarised signal-to-noise ($\mathcal{L}=pI/\sigma_{pI}$). We show this distribution in Figure~\ref{fig:rm_polint}. Similar to the distribution we saw in \citetalias{Thomson2023}, the distribution of RM continues smoothly below the nominal cut of $8\sigma$ in $pI$. In this case, however, we see that the apparently smooth RM distribution continues down to $\sim5\sigma$ before abruptly spanning the entire possible range of RMs. In \citetalias{Thomson2023}, this occurred at $\sim7\sigma$ (see Figure~8 of that work). Here we can also see the effect of our two channelisations, namely 36 or 72 channels, in the sharp cuts at RM$\sim\qty{600}{\radian\per\metre\squared}$ and $\sim\qty{1200}{\radian\per\metre\squared}$.

\begin{figure}
    \includegraphics{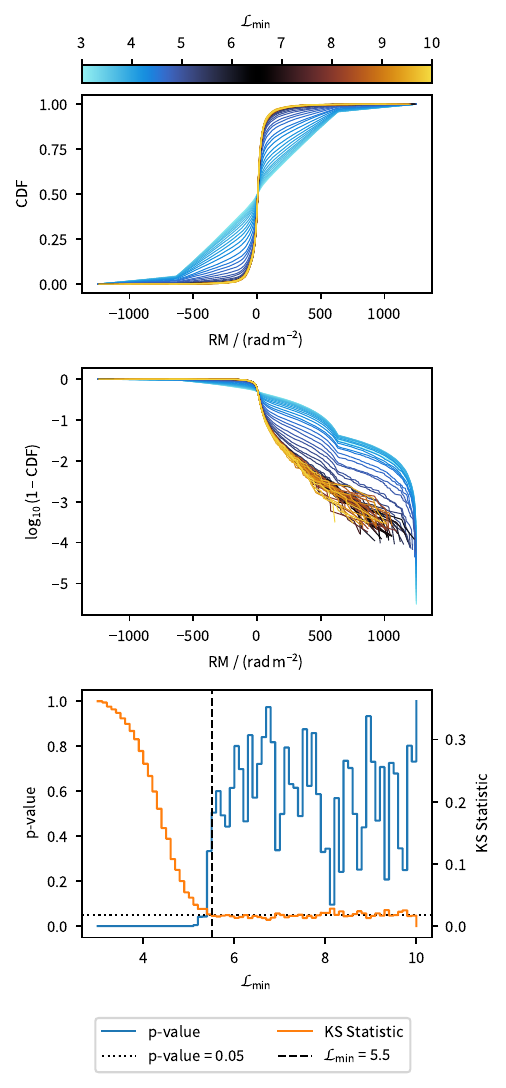}
    \caption{
        A comparison of distributions in polarised signal-to-noise ($\mathcal{L}=pI/\sigma_{pI}$) bins. Here the value $\mathcal{L}_\text{min}$ gives the left (inclusive) edge of a bin, with a width of 0.1. The upper and middle panels show the cumulative distribution function (CDF) of RM in each bin, where the colour of each line corresponds to the $\mathcal{L}_\text{min}$ bin. The middle panel rescaled to highlight small separations between the CDFs at large (postive) RMs. The bottom panel shows the results of a two-sample Kolmogorov-Smirnov (KS) test between each bin and the highest signal-to-noise bin. In orange we show the value of the KS statistic, and in blue we show the associated p-values. The horizontal, dotted line shows where the p-value is 0.05, and the vertical, dashed line shows the lowest $\mathcal{L}_\text{min}$ bin for which we cannot reject the null-hypothesis (see text in \S\ref{sec:rm-grid} for further details).
    }
    \label{fig:cdfs}
\end{figure}

To investigate where this unphysical distribution of RMs occurs, we slice our catalogue into subsets of $\mathcal{L}$ values. We form bins with a lower edge $\mathcal{L}_\text{min}$ and a width $\delta_{\mathcal{L}}$ such that a bin is defined by $\mathcal{L}_\text{min} \leq \mathcal{L} < \mathcal{L} + \delta_{\mathcal{L}}$. We chose lower edges in the range $3\leq \mathcal{L}_\text{min} \leq 10$ and take $\delta_{\mathcal{L}}=0.1$. In the upper panels of Figure~\ref{fig:cdfs} we show the cumulative distribution function (CDF) of RM for each signal-to-noise ratio. To compare the distributions we take a two-sample Kolmogorov-Smirnov (KS) test between the highest $\mathcal{L}$ bin and all other bins. In the bottom panel of Figure~\ref{fig:cdfs} we show both the KS statistic as well as the p-value corresponding to the null hypothesis. In this case the null hypothesis is that the two samples were drawn from the same distribution; if the p-value is below a selected threshold we can reject the null hypothesis and conclude that the samples were drawn from different distributions. If we take a threshold p-value of 0.05, corresponding to a confidence of \qty{95}{\percent}, we find that for $\mathcal{L}_\text{min}$ below $5.5\sigma$ we can reject null the hypothesis. Stated the other way, for each bin where $\mathcal{L}_\text{min}\geq 5.5$ we cannot discriminate between the distribution of RM in that bin and the distribution in the $10\sigma$ bin.

A robust analytic formalism for detection thresholds in linear polarisation and RM-synthesis was derived by \citet{Hales2012}, with similar discussions in \citet{Macquart2012} and \citet{George2012}. Equation 31 of \citet{Hales2012} provides the required value of $\mathcal{L}$ for an equivalent false detection in a Gaussian distribution, with signal-to-noise ratio ($\mathcal{G}$):
\begin{equation}\label{eqn:ell}
    \mathcal{L} = \sqrt{-2 \ln\left\{1 - \left[\text{erf}\left(\frac{\mathcal{G}}{\sqrt{2}}\right)\right]^{1/N_\text{FDF}}\right\}},
\end{equation}
and inversely \citep[][Equation 30]{Hales2012},
\begin{equation}\label{eqn:gee}
    \mathcal{G}=\sqrt{2}\text{erf}^{-1}\left\{\left[1-\exp{\left(\frac{-\mathcal{L}^2}{2}\right)}\right]^{N_\text{FDF}}\right\}.
\end{equation}
Here $\text{erf}$ and $\text{erf}^{-1}$ are the error function and its inverse, respectively, and $N_\text{FDF}$ is the number of independent samples across the FDF given by:
\begin{equation}\label{eqn:n_fdf}
    N_\text{FDF} = \frac{2\phi_\text{max}}{\delta\phi} \approx \frac{\Delta\lambda^2}{\delta\lambda^2}.
\end{equation}
Here, $\phi_\text{max}$ and $\delta\phi$ are as defined in \S\ref{sec:imaging}. We evaluate $N_\text{FDF}$, $\mathcal{L}$, and $\mathcal{G}$ for both of our two channelisations, as well as the 288 channels used by POSSUM and the lower limit of 2 channels; which we summarise in Table~\ref{tab:detection}. By construction, for the same bandwidth, $N_\text{FDF}$ scales proportionally to the number of frequency channels. As such, the false detection likelihood is higher when using finer channelisation. For the channelisations we use here, the polarised SNR required for a Gaussian equivalent of $\mathcal{G}=5$ is $\mathcal{L}\sim6$. Conversely, the value we discussed above of $\mathcal{L}=5.5$ corresponds to $\mathcal{G}\sim4.5$, with a false detection rate of 1 in $1.7 \times 10^{5}$ and 1 in $8.2 \times 10^{4}$ for 36 and 72 channels, respectively. 

It is likely a polarised signal-to-noise cut of $5.5$ may begin to include false positive detections, but for our sample a $6\sigma$ cut should be equivalent to a $5\sigma$ limit of a Gaussian distribution. Using such a limit yields 466\,111 total RMs, with 342\,606 after de-duplication. As we mention in \S\ref{sec:subsets}, however, it is more important to properly weight the RMs in a grid, than to find the ideal strict cutoffs \citep{Rudnick2019}. For the remainder of our analysis here we will continue with the \texttt{goodRM} cut of $8\sigma$ as a conservative limit.

\begin{table}
\begin{tabular}{ccccccc}
\toprule
$N_\text{chan}$ & $N_\text{FDF}$ & \multicolumn{2}{c}{$\mathcal{L}$} & \multicolumn{3}{c}{$\mathcal{G}$} \\
\cmidrule(r){3-4}\cmidrule(r){5-7}
& & $\mathcal{G}=5$ & $\mathcal{G}=8$ & $\mathcal{L}=5$ & $\mathcal{L}=5.5$ & $\mathcal{L}=8$ \\
\midrule
2 & 1 & 5.4 & 8.3 & 4.6 & 5.1 & 7.7 \\
36 & 22 & 5.9 & 8.6 & 3.9 & 4.5 & 7.3 \\
72 & 45 & 6.0 & 8.7 & 3.8 & 4.4 & 7.2 \\
288 & 183 & 6.3 & 8.9 & 3.4 & 4.1 & 7.0 \\
\bottomrule
\end{tabular}
\caption{
    Detection statistics for different channelisations ($N_\text{chan}$). Here $N_\text{FDF}$ is the number of independent samples across the FDF given our frequency coverage, assuming an idealised, uniformly-weighted RMSF. The third and fourth columns both provide the linearly polarised signal-to-noise ratio ($\mathcal{L}=pI/\sigma_{pI}$) for a given Gaussian-equivalent signal-to-noise ratio ($\mathcal{G}$). In the fifth to seventh column we provide the inverse; the values of $\mathcal{G}$ for a given $\mathcal{L}$.  We use derivation and formalism of these values from \citet{Hales2012} as given in Equations~\ref{eqn:ell}, \ref{eqn:gee}, and \ref{eqn:n_fdf}.
}\label{tab:detection}
\end{table}

\begin{figure}
    \includegraphics[width=\columnwidth]{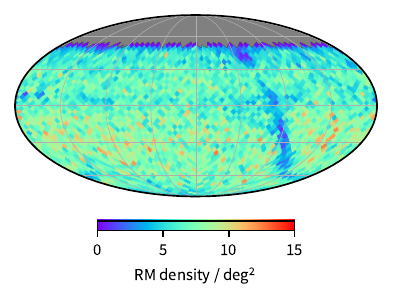}
    \caption{
        The areal density of components with well-determined RMs in SPICE-RACS DR2 after de-duplication. The density of components is calculated on a HEALPix grid with $N_\text{side}=16$, corresponding to a pixel resolution of $\sim\ang{;220}$, in celestial coordinates.
    }
    \label{fig:rm_density}
\end{figure}

Taking the \texttt{goodRM} subset of our de-duplicated catalogue we can look at the areal density of RMs across our survey area. In Figure~\ref{fig:rm_density} we show the density of RMs in a HEALPix grid across the sky. Using this pixelisation, we find a median and IQR density of $6.7 \substack{+1.8\\-1.7}\unit{\perdegreesq}$; an improvement of $\sim\qty{75}{\percent}$ over DR1~\citepalias{Thomson2023}. The spatial variance in RM density anti-correlates (Spearman's $r=-0.6$) as expected with the estimated $rms$ noise. Along the Galactic plane within the inner Galaxy we see elevated $rms$, however, we also expect to see elevated RM complexity \citep{Livingston2021} and therefore increased depolarisation. We will discuss our own Farday complexity metrics in \S\ref{sec:complex}. Overall, however, towards the inner Galaxy it is difficult to disentangle the impact of increased $rms$ against Faraday complexity on the RM density using our dataset alone. We note that this region is simultaneously astrophysically complex and technically difficult to image using a snapshot survey such as RACS. As such, users of our data will need to take particular care in the analysis of our data products in this region.

\begin{landscape}
    \begin{figure}
        \centering
        \includegraphics{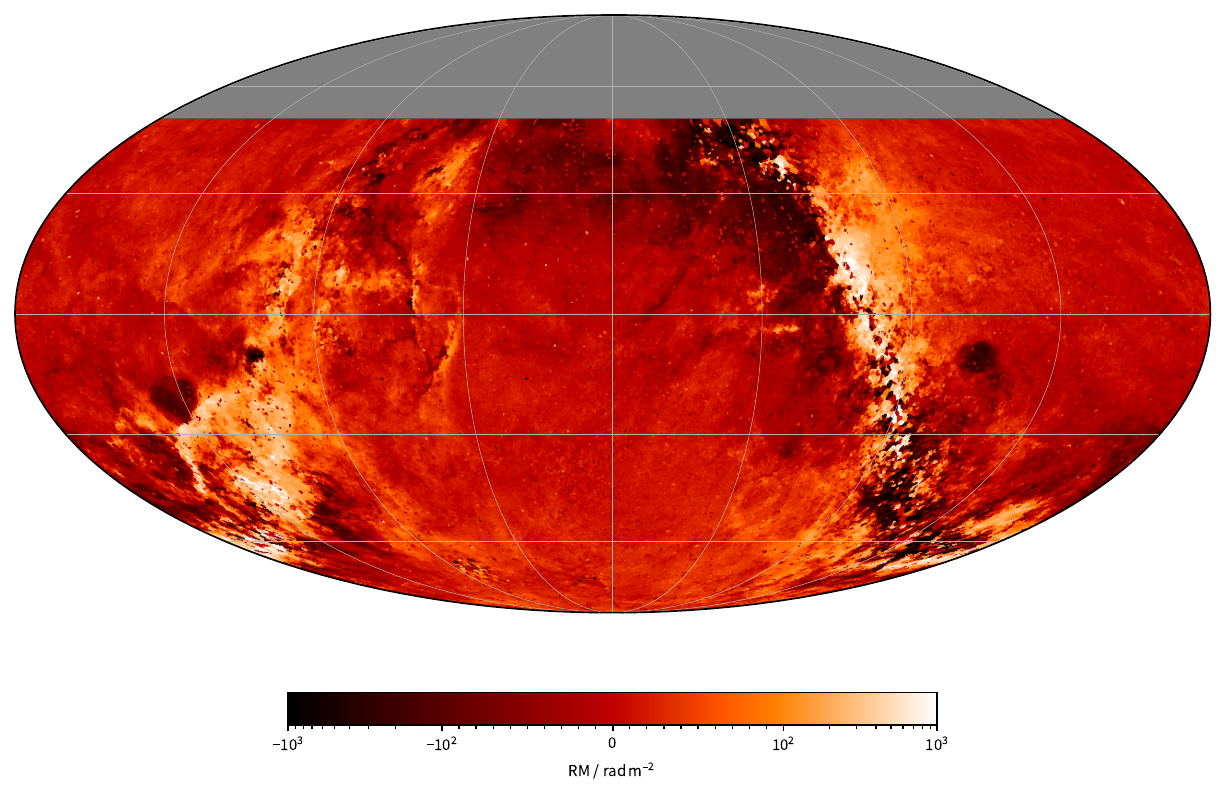}
        \caption{Rotation measures (RM) across the survey area in celestial coordinates using nearest-neighbour interpolation. Here we use a colourmap with a symmetric logarithmic scale. Values with $|\text{RM}|\leq \qty{100}{\radian\per\meter\squared}$ are shown on a linear scale, and values above this range are logarithmically scaled. We note that this image is not of directly detected diffuse emission, rather it is a visualisation of our catalogue data. We provide a linearly-scaled version of this image in Galactic coordinates in Figure~\ref{fig:rm_linear}.}
        \label{fig:rm_non_linear}
    \end{figure}

\end{landscape}

We show the RM grid from our de-duplicated catalogue in Figure~\ref{fig:rm_non_linear} using a nearest-neighbour interpolation. We show an alternate presentation of this image in the Appendix Figure~\ref{fig:rm_linear}. Given the average density of $\sim\qty{7}{\perdegreesq}$, our RM grid has an effective `resolution' of \ang{;23}. Our broadband RMs also give excellent RM precision; the median and IQR of the RM error is $2.2 \substack{+1.1\\-1.3}\unit{\radian\per\metre\squared}$.

Many of the familiar large-scale features of the Faraday sky~\citep[e.g.][]{Hutschenreuter2021} emerge from this image. On smaller scales, our improved RM density has now resolved a plethora of Galactic features. Of particular note are striking filamentary RM structures, many of which appear to be very linear as projected on the sky. Many of these features are well studied objects, such as large scale H$\alpha$ bubbles and filaments \citep[e.g.]{Haffner1998a, Madsen2006, Stil2016, Hutschenreuter2024}. However, our data reveal features that extend completely across both the Northern and Southern Galactic poles. These bear resemblance to similar structures seen in neutral interstellar medium tracers such as \textsc{Hi} or interstellar dust~\citep[see][and references therein]{McClureGriffiths2023}.

\begin{figure*}
    \centering
    \begin{subfigure}{\textwidth}
        \includegraphics[width=\textwidth]{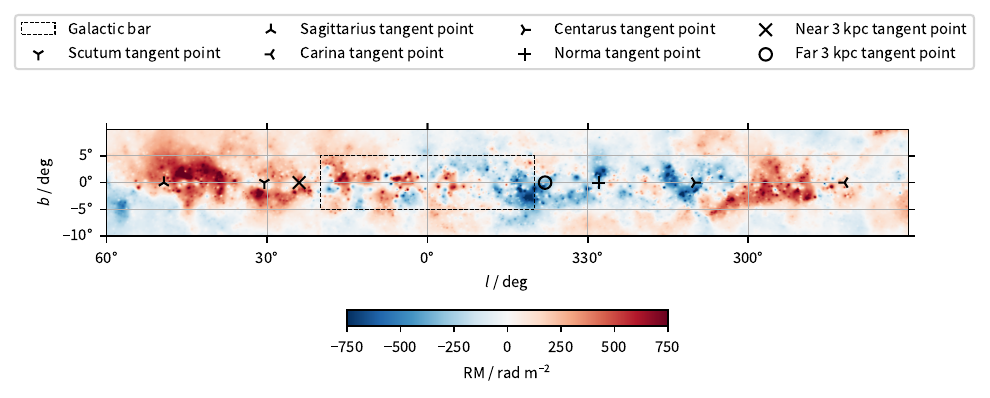}
        \caption{Inner Milky Way.}\label{subfig:cool_rm_galcen}
    \end{subfigure}
    \begin{subfigure}{0.49\textwidth}
        \includegraphics[width=\textwidth]{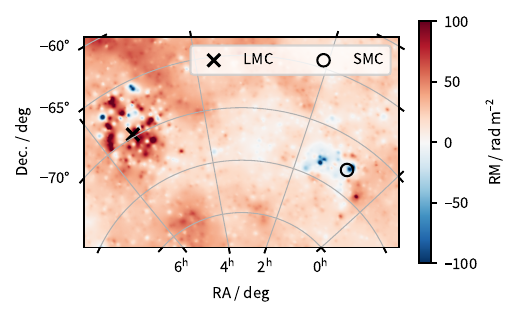}
        \caption{Large (LMC) and Small (SMC) Magellanic Clouds.}\label{subfig:cool_rm_mc}
    \end{subfigure}
    \begin{subfigure}{0.49\textwidth}
        \includegraphics[width=\textwidth]{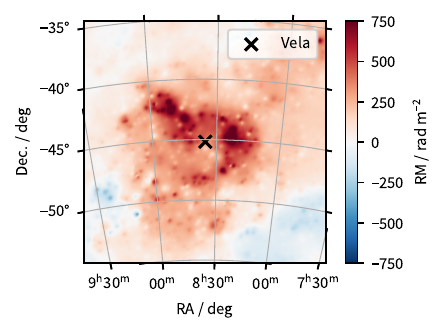}
        \caption{Vela supernova remnant.}\label{subfig:cool_rm_vela}
    \end{subfigure}
    \begin{subfigure}{0.49\textwidth}
        \includegraphics[width=\textwidth]{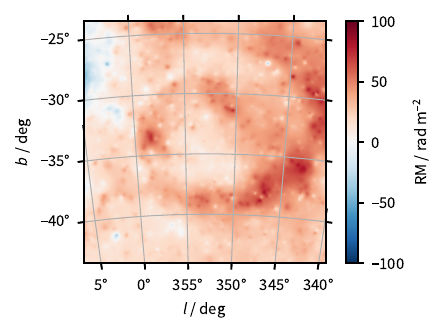}
        \caption{G353-34.}\label{subfig:cool_rm_353}
    \end{subfigure}
    \caption{
        A selection rotation measure (RM) structures in SPICE-RACS DR2. We discuss each of these features in \S\ref{sec:rm-grid}. In each panel we show the RM sky using linear interpolation with inverse distance-squared weighting. We note that these images are not of directly detected diffuse emission, rather they are a visualisation of our catalogue data.
    }
    \label{fig:cool_things}
\end{figure*}

We will briefly remark on a select number of notable individual features, which we show in Figure~\ref{fig:cool_things}. In Figure~\ref{subfig:cool_rm_galcen} we show the inner Galactic plane between longitudes $l=\ang{60}$ and $l=\ang{270}$. In the dashed box we show the approximate location of the Galactic bar~\citep{Churchwell2009}. Models of the bar place the portion in the first quadrant ($\ang{0}<l<\ang{90}$) as the section closer to the Sun. In this section we see a quasi-periodic RM pattern, with a period of $\sim\ang{10}$, with antisymmetry about the Galactic plane, forming a `zipper-like' pattern. The reversals in RM sign must correspond to line of sight reversals in the Galactic magnetic field. Further analysis is warranted to investigate whether this is a feature of the Galactic centre / bar, or whether it caused by some foreground screen on the Galactic plane.

In this same figure we also overlay the approximate tangent points for a number of spiral arms from \citet{Hou2014}. Towards the Scutum and Norma arms we see local enhancements in the RM, with magnetic fields pointing towards and away from the Sun, respectively. The enhancements are also offset South and North of the plane for the Scutum and Norma arms, respectively. Towards the Sagittarius arms we see a clear continuation of the RM reversal reported by \citet{Ordog2017}. 

In Figure~\ref{subfig:cool_rm_mc} we see the expected RM enhancement towards the Magellanic clouds as previously reported by \citet{Mao2008,Mao2012,Livingston2022,Livingston2024}. For the first time, however, we have a uniformly dense RM grid over the Magellanic clouds, Bridge, and Stream. A detailed investigation of this region with our catalogue is forthcoming in McClure-Griffiths et al. (in prep.).

Towards the Vela supernova remnant we see a significant RM enhancement with complex spatial structure. We see a ring-like enhancement in positive RM which follows the Eastern edge of the supernova remnant as seen in radio continuum~\citep[e.g.][]{Calabretta2014}. Towards the North East of the region we see a more linear feature, running from the edge of the ring towards the North East. A similar prominence is seen in radio continuum, and is suggestive of chimney-like structure where the RM could be enhanced by an over-density of thermal electrons.

Finally, we remark on a unique large-scale feature of the Southern RM sky G353-34. This large ring-like structure was first discovered by \citet{Testori2008} in diffuse polarisation images at \qty{1.4}{\giga\hertz} as a depolarisation feature. The precise nature of the object has not been determined, however it is assumed to be a nearby supernova remnant based on its morphology. G353-34 is also visible in S-PASS~\citep{Carretti2019} as polarisation angle feature at \qty{2.3}{\giga\hertz}, and in both depolarisation and RM in STAPS~\citep{Raycheva2025,Sun2025}. In our RM grid we see G353-34 as a large ring of enhanced positive RM.

This is far from a comprehensive summary of all of the features and structures present in our RM grid. In particular, discovery and analysis of extragalactic RM features will require careful modelling and subtraction of the Milky Way foreground. 

\subsubsection{Galactic foreground RM}\label{sec:nn_foreground}
To provide users of our catalogue with a first order estimate of the Galactic contribution, we make use of the nearest neighbour average method described in \citet{Anderson2024} and Malik et al. (in prep.). For every single unique position in our catalogue we find a set of nearest-neighbouring sources in the deduplicated \texttt{goodRM} subset, and compute a number of statistics on that set. Within each set also we exclude components that are within three major beam widths of the reference positions. Here we select a value of 35 neighbours. We found this gave a good balance between the number of samples on average and the mean maximum angular separation of components ($\ang{1.3}\pm\ang{0.2})$.

For each of neighbouring polarised components we compute the following statistics:
\begin{itemize}
    \item Number of nearest neighbours.
    \item The median RM.
    \item The mean RM.
    \item The standard deviation of the RMs.
    \item The median absolute deviation (MAD) of the RMs scaled to the standard deviation.
    \item The standard error of the mean of the RMs.
    \item The error-weighted mean RM.
    \item The standard error of the mean of the error-weighted mean RM.
    \item The maximum angular separation between the components.
    \item The minimum angular separation between the components.
\end{itemize}
We provide the corresponding column names for these values in \S\ref{sec:catalogue}.

We caution against direct use of these statistics as the `true' foreground RM from the Milky Way. The angular scales of the Milky Way foreground RM is highly direction dependent, and may overlap a given background area and scale of scientific interest. Instead, we provide these statistics as a first-pass indication of how complex a given area may be with respect to the foreground RM structure. In particular, users of our catalogue should be very cautious of any foreground estimate where our standard deviation statistics are high.

\begin{figure}
    \centering
    \begin{subfigure}{\textwidth}
        \includegraphics[width=\textwidth]{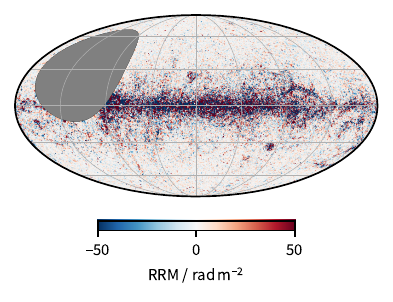}
        \caption{}\label{subfig:grm}
    \end{subfigure}
    \begin{subfigure}{\textwidth}
        \includegraphics[width=\textwidth]{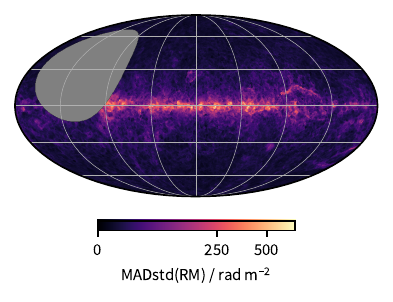}
        \caption{}\label{subfig:grm_err}
    \end{subfigure}
    \caption{
         Interpolated sky maps resulting from our nearest-neighbour foreground RM estimates. In (\subref{subfig:grm}) we show the residual RM (RRM) having subtracted median RM of the ensemble of neighbours from each RM. In (\subref{subfig:grm_err}) we show the median absolute deviation scaled to the standard deviation (MADstd) of the RM in the ensemble of neighbours. Both maps are shown in Galactic coordinates centred on $l,b=(\ang{0},\ang{0})$.
    }
    \label{fig:grm}
\end{figure}

We show the resulting RRM and the MADstd of the RM from this foreground analysis in Figure~\ref{fig:grm}. Regions of high absolute RRM correlate as expected with high MADstd(RM). Further, the resulting MADstd(RM) image strongly resembles all sky images tracing emission measure, such as H$\alpha$~\citep[e.g.]{Reynolds1998,Finkbeiner2003}. These features extend even to Galatic latitudes as high as high as $|b|\sim60\degr$. Conversely, there are also a number of `windows' which represent sight-lines with much reduced Galactic foreground contamination. These include areas towards the Galactic North pole, and a region in the Galactic South within $-30\degr\lesssim l\lesssim30\degr$. A deep analysis of the spatial statistics of these regions, and the entire SPICE-RACS DR2 area, is forthmcoming in Anderson et al. (in prep.).

\subsection{Faraday complexity}\label{sec:complex}
Our broad bandwidth affords us the ability to characterise Faraday complexity; where multiple components are detected in the FDF. It is important to note that our FDF resolution $\delta\phi\approx\qty{63}{\radian\per\metre\squared}$ is larger than the maximum scale $\phi_\text{max-scale}\approx41$. As such, Faraday thick structures will be depolarised, or `resolved out', in Faraday depth space.

As discussed in further detail in \citetalias{Thomson2023}, we make use of the two complexity metrics that \textsc{RM-Tools} provides with some simple modifications. The first is $\sigma_\text{add}$, which characterises the additional variance present in a spectrum after a Faraday simple model is subtracted from the peak RM. The second is $m_2$ which is the second moment of the Faraday \textsc{clean} components, which we normalise by $\delta\phi$. We flag a spectrum as complex if $\sigma_\text{add} >1$ and $\sigma_\text{add}/\sigma_{\sigma_\text{add}}>10$ (where $\sigma_{\sigma_\text{add}}$ is the error on $\sigma_\text{add}$), or if $m_2>1$. We summarise the counts and fraction of components with our complexity flags in Table~\ref{tab:complex}.

Overall, only \qty{1.4}{\percent} of our components in the \texttt{goodRM} subset are classified as complex. By way of comparison, in DR1 the fraction was \qty{12}{\percent}. We will return to this discrepancy below. Again we note that the $\sigma_\text{add}$ metric is a more sensitive identifier; $\sigma_\text{add}$ flags an order of magnitude more components as complex against $m_2$. Only 849 components are classed as complex by both metrics.

\begin{table}
\centering
\begin{tabular}{ccccc}
\toprule
\multicolumn{3}{c}{Complexity flag} & Count & Fraction  \\
\cmidrule(r){1-3}
$m_2$ & $\sigma_\text{add}$ & $m_2$ or $\sigma_\text{add}$  & & \% \\
\midrule
False & False & False & 328657 & 98.6 \\
False & True & True & 3188 & 0.957 \\
True & False & True & 479 & 0.144 \\
True & True & True & 849 & 0.255 \\
\midrule
--  & --  &  True  & 4516    & 1.36 \\
\bottomrule
\end{tabular}
\caption{
    Counts of components flagged as Faraday complex by each of our metrics, $m_2$ and $\sigma_\text{add}$. These flags can found in our catalogue in the \texttt{complex\_M2\_CC\_flag}, \texttt{complex\_sigma\_add\_flag}, and \texttt{complex\_flag} columns.
}\label{tab:complex}
\end{table}

\begin{figure}
    \centering
    \includegraphics[width=\linewidth]{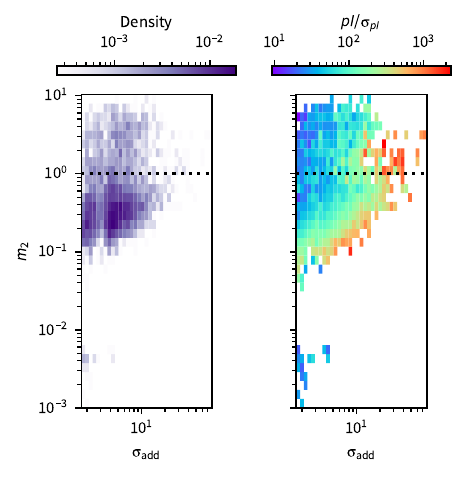}
    \caption{
        A comparison of our Faraday complexity metrics, $m_2$ and $\sigma_\text{add}$ in a 2D histogram. In the left panel we show the density of components from the \texttt{goodRM} subset in each bin. In the right panel we show the median polarised signal-to-noise $pI/\sigma_{pI}$ in each bin. In the black, dashed line we show where $m_2=1$; components where $m_2>1$ are classed as complex by this metric. Note that all components shown here are classified as complex by $\sigma_\text{add}$.
    }
    \label{fig:complex_complex}
\end{figure}

Having inspected the complexity metrics, we find a few dimensions that correlate with these values. In Figure~\ref{fig:complex_complex} we show the 2-dimensional histogram of $m_2$ against $\sigma_\text{add}$, and median value of $pI/\sigma_{pI}$ in the same bins. Again similar to \citetalias{Thomson2023} we see a correlation of complexity with $pI/\sigma_{pI}$; particularly so for $\sigma_\text{add}$. As discussed in \citet{Alger2021}, complexity metrics are not necessarily looking for `complex' spectra, rather they identify spectra which are `not simple'. That is to say, in the low signal-to-noise regime we cannot tell a simple from a complex source. This naturally biases detections of complex sources to high values of $pI/\sigma_{pI}$.

\begin{figure}
    \centering
    \includegraphics[width=\columnwidth]{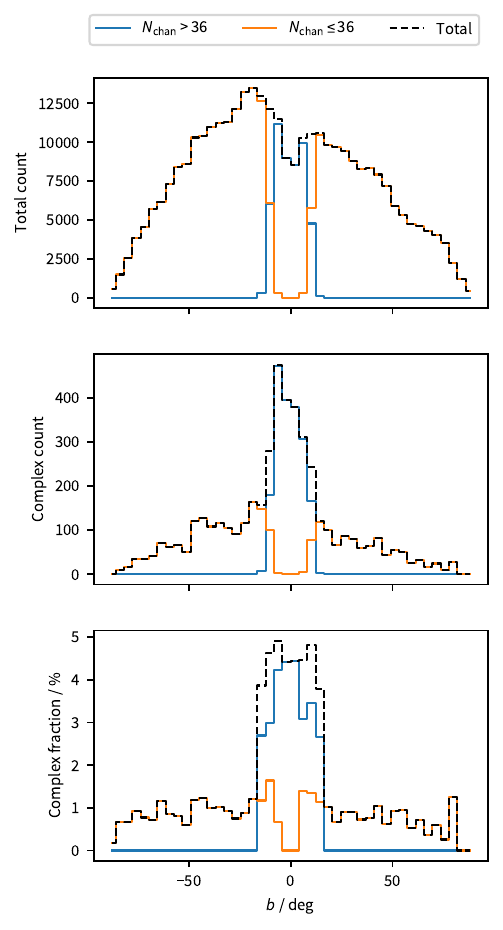}
    \caption{
        Counts of components in our \texttt{goodRM} subset in bins of Galactic latitude $b$. For all panels we show total counts in black, and counts where the number of spectral channels ($N_\text{chan}$) is $>36$ or $\leq36$ in blue and orange, respectively. In the top panel we show the counts for all components, in the middle we show counts of components flagged as Faraday complex, and in the bottom we show the fraction of components flagged as complex.
    }
    \label{fig:complex_counts}
\end{figure}

In addition to $pI/\sigma_{pI}$, as shown in Figure~\ref{fig:complex_counts}, we see a correlation of the number of complex components with Galactic latitude. However, Galactic latitude also defines the primary heterogeneity in our data processing; the number of spectral channels. In Figure~\ref{fig:complex_counts} we also show the break down of components by the number of spectral channels (either $>36$ or $\leq36$). The sharp jump in the fraction of complex components towards the Galactic plane is clearly caused by the transition from our selection of 36 or 72 imaging channels. We do have overlapping fields with different numbers of channels that cover the same areas of sky. Where the number of channels $\leq36$ we see that the number of complex components is approximately constant ($\sim\qty{1}{\percent}$) for all latitudes sampled. For components where the number of channels is $>36$ we see some indication that the fraction of complex components does increase as Galactic latitude decreases. We refrain from making a definitive conclusion in this case as the number counts of complex components is relatively small, which necessitates wide bins in latitude.

In \citetalias{Thomson2023} we discussed the known \qty{25}{\mega\hertz} spectral ripple in ASKAP observations, and its potential impact on Faraday complexity metrics. This feature is still present in RACS-low3 observations; however, here we have changed our nominal frequency coverage and channelisation with respect to SPICE-RACS DR1. 

\begin{figure}
    \centering
    \includegraphics[width=\linewidth]{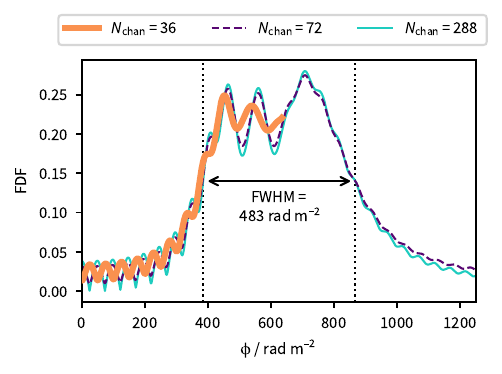}
    \caption{
        A model FDF for a spectral ripple with a \qty{25}{\mega\hertz} period. We show the FDF as produced by different channelisations ($N_\text{chan}$) of the RACS-low3 band. We indicate the full-width at half-maximum (FWHM) of the FDF. We note that the FDF is symmetric about $\phi=0$, and here we are just showing the FDF where $\phi>0$.
    }
    \label{fig:ripple}
\end{figure}

To model such a ripple in our band, we simulate a sinusoidal oscillation in Stokes $Q$ and $U$ with a \qty{25}{\mega\hertz} period and a unit amplitude. We sample with model with 36, 72, and 288 channels. We show the resulting FDFs after RM-synthesis in Figure~\ref{fig:ripple}. We find that a \qty{25}{\mega\hertz} ripple creates a symmetric feature about $\phi=0$ with a peak of about \qty{28}{\percent} of the input amplitude. This feature is broad in Faraday depth with full-width at half-maximum $\text{FWHM}=\qty{483}{\radian\per\metre\squared}$. The half-power points of this feature are at $\phi=\pm\qty{385}{\radian\per\metre\squared}$ and $\pm\qty{868}{\radian\per\metre\squared}$. 

We see minimal difference in the appearance of the feature between channelisations of 72 or 288. Notably, in the case of 36 channels the cut of $\phi_\text{max}$ ends the sampling of the ripple feature about half way between its half power points, which is also before the peak in the FDF. The amplitude scale is also marginally reduced. We note that using 36 channels provides us a sampling interval of \qty{8}{\mega\hertz}, which is less than the Nyquist interval of \qty{12.5}{\mega\hertz} for the spectral ripple. The ripple is therefore oversampled in all cases. However,  as discussed in \S\ref{sec:imaging}, the beaviour of \textsc{RM-Tools} is to select maximum Faraday depth by taking the largest $\lambda^2$ channel spacing after converting from frequency spacing. As such our data with only 36 channels only samples out to $|\phi|\lesssim\qty{630}{\radian\per\metre\squared}$ and does not sample the full range of ripple feature in Faraday depth space.

Overall, we suggest that the spectral ripple is likely contributing to our measured complexity towards the Galactic plane. Where we use 36 channels our smaller maximum Faraday depth range likely limits the impact of this ripple. It is also critical to note that, as per \citetalias{Thomson2023}, the Faraday depth of a ripple feature can either remained fixed or translate with the source RM in Faraday depth. The specific behaviour depnds on whether the ripple is introduced to Stokes $Q$ and $U$ via leakage from Stokes $I$ or if it is intrinsic to the antenna gains. On this basis we conclude that the higher fraction on complex components observed in \citetalias{Thomson2023} was due to the higher likely impact of the residual bandpass ripple. Further investigation of intrinsic Faraday complexity will likely require the addition of higher frequency observations (as discussed in \S\ref{sec:conclusion}), and improved suppression of systematics.

\subsection{Time-domain analysis}\label{sec:time}

In this data release we have retained the time-domain information from RACS-low3 by simply concatenating our per-field catalogues. For the purposes of this work, we use this time domain information to quantify the reliability of our RM catalogue. We leave variability analysis and other time-dependent analysis to future work.

\begin{figure}
    \centering
    \begin{subfigure}{\columnwidth}
        \includegraphics[width=\textwidth]{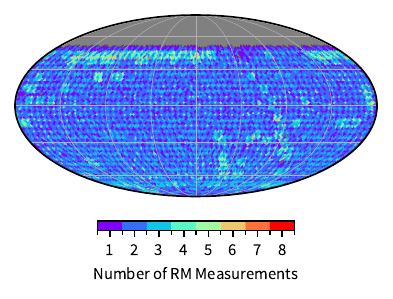}
        \caption{}\label{subfig:repeat_count}
    \end{subfigure}
    \begin{subfigure}{\columnwidth}
        \includegraphics[width=\textwidth]{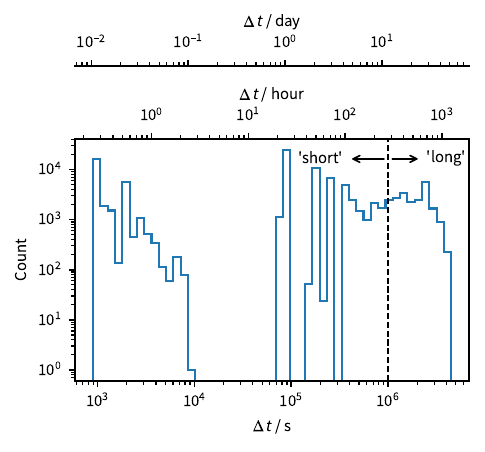}
        \caption{}\label{subfig:time_diff}
    \end{subfigure}
    \caption{
        Time-domain sampling in our \texttt{goodRM} subset. In (\subref{subfig:repeat_count}) we show the number of repeated observations of each component across the sky. In (\subref{subfig:time_diff}) we show the probability density (PDF) of the separation in time ($\Delta t$) between observations of each component. Given the large range of time samples, we provide scales in seconds, hours, and days. We define `short' and `long' subsets as $\Delta t$ being less than or greater than \qty{1e6}{\second}, respectively.
    }\label{fig:time_sky}
\end{figure}

In Figure~\ref{subfig:repeat_count} we show the number of repeated observations of each component in the \texttt{goodRM} subset across the survey area. At most we have 8 observations of the same component. Here we restrict our analysis primarily to the apparent variability of RM. As such, we exclude all components marked as Faraday complex to remove the effect of selecting different peaks. For each component with repeated observations we aggregate our catalogue on components, and compute the error-normalised difference in RM between pairs of observations using Equation~\ref{eqn:error_diff}.

RACS-low3 was not designed as a self-contained time-domain survey, and as such our $\Delta t$ is highly non-uniform; we show this distribution in Figure~\ref{subfig:time_diff}. Our shortest time spacing is \qty{895}{\second}, as short as reasonably possible with \qty{15}{\minute} integrations, and our longest is \qty{52}{\day}. We note that we particularly lack sampling for $\qty{3}{\hour}\lesssim\Delta t\lesssim\qty{16}{\hour}$. For later analysis we define two subsets based on the time spacing. We define the subset where $\Delta t\leq$\qty{1e6}{\second} ($\sim\qty{11.6}{\day}$) as `short' and the remainder as `long'.

\begin{figure}
    \includegraphics{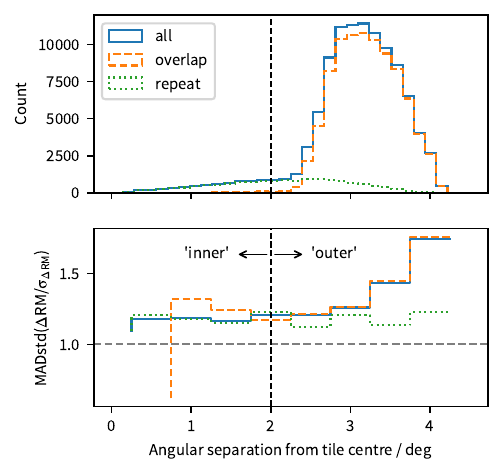}
    \caption{
        Pairs of RM measurements across repeated observations in SPICE-RACS DR2. Upper panel: Number count histogram for component pairs as a function of separation from a given tile centre. Lower panel: MADstd of error-normalised RM difference between pairs in bins of angular separation from a given tile centre. In blue, solid lines we show all component pairs, in orange, dashed we show components from overlapping tiles, and in green, dotted lines we show components from repeated tile observations. We define `inner' and `outer' subsets as being inside or outside of \ang{2} from a tile centre, respectively.
    }\label{fig:space_sky}
\end{figure}

Repeated observation in RACS occur from either overlapping tile edges or from repeated tile observations. In the former case, there is a strong bias to select components towards the edge of each tile. Regions on tile edges have intrinsically higher systematics, such as polarisation leakage, and our uncertainty on corrections to such systematics are also higher. In the top panel of Figure~\ref{fig:space_sky} we show the number count distribution of RM pairs from both overlapping and repeated tiles. We see that the vast majority of repeated observations (\qty{73.8}{\percent}) are from overlapping observations.

We now look at the distribution of $\Delta\text{RM}/\sigma_{\Delta\text{RM}}$ across our repeated pairs and its subsets. In the lower panel of Figure~\ref{fig:space_sky} we show the $\text{MADstd}(\Delta\text{RM}/\sigma_{\Delta\text{RM}})$ as a function of separation from a given tile centre. If the error distribution in RM were ideally Gaussian we would expect $\Delta\text{RM}/\sigma_{\Delta\text{RM}}$ to follow a standard normal distribution. We see that repeated observations have a MADstd close to 1 across all separations. For the overlapping fields, however, we see that variance increases towards larger angular separations.

\begin{figure}
    \centering
    \begin{subfigure}{\columnwidth}
        \includegraphics[width=\textwidth]{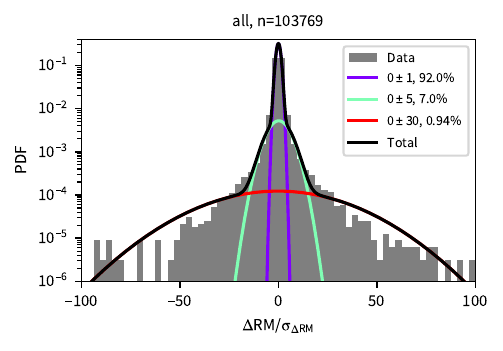}
        \caption{}\label{subfig:rm_diff_hist_all}
    \end{subfigure}
    \begin{subfigure}{\columnwidth}
        \includegraphics[width=\textwidth]{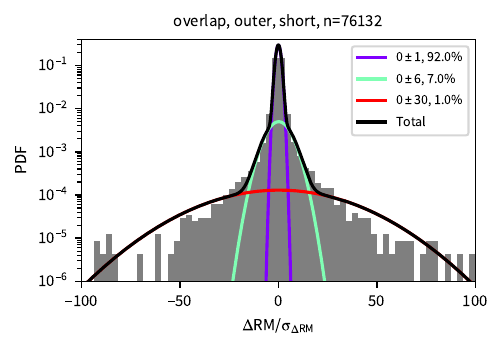}
        \caption{}\label{subfig:rm_diff_hist_overlap_outer_short}
    \end{subfigure}
    \begin{subfigure}{\columnwidth}
        \includegraphics[width=\textwidth]{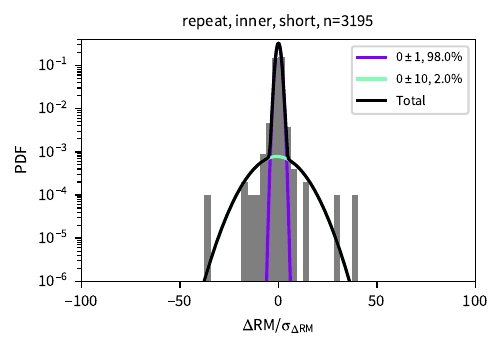}
        \caption{}\label{subfig:rm_diff_hist_repeat_inner_short}
    \end{subfigure}
    \caption{
        The distributions of RM changes across time for Faraday simple components. In (\subref{subfig:rm_diff_hist_all}) we show the probability density of error-normalised RM differences for all pairs of repeated observations. In (\subref{subfig:rm_diff_hist_overlap_outer_short}) and (\subref{subfig:rm_diff_hist_repeat_inner_short}) we show the same except for the `outer', `short' subset of overlapping tiles and the `inner', `short' subset of repeated tiles, respectively.  In the solid black line we show our fitted Gaussian Mixture Model (GMM), with each component of the model coloured separately. We give the number count ($n$) of each subset in the title of each panel. Additionally, we give the mean, standard deviation, and weight of each GMM component in the respective legends.
    }\label{fig:rm_diff_hist}
\end{figure}

Inspecting the distribution of $\Delta\text{RM}/\sigma_{\Delta\text{RM}}$ directly, as show in Figure~\ref{subfig:rm_diff_hist_all}, we see a multi-component distribution. We note a clear standard normal distribution, with a number of broader distributions and outliers superposed. To quantify the fraction of data within and without of a standard normal distribution we make use of a Gaussian Mixture Model (GMM). Specifically, we use \pkg{sklearn.mixture.GaussianMixture} from \textsc{Scikit-learn}~\citep{scikit-learn} to both estimate the number of required components and to fit to the data. The initial fit to the entire dataset shows that \qty{92}{\percent} of components are consistent with a standard normal distribution. As noted above, however, the overall set is heavily biased to sampling components on the edge of individual tiles. As such, we look to our subsets.

We first turn our attention to the subset of component pairs we expect to have the highest level of residual systematics; the outer region of overlapping tiles. We also select on our `short' time scale to give us confidence that any variance in $\Delta\text{RM}/\sigma_{\Delta\text{RM}}$ is likely not from an astrophysical source. We show this distribution in Figure~\ref{subfig:rm_diff_hist_overlap_outer_short}. We see a nearly identical result to the overall set, with \qty{92}{\percent} of components being fit by a standard normal distribution. The GMM process also fits two 0-mean components with standard deviations of 6 and 30, with weights of \qty{7}{\percent} and \qty{1}{\percent}, respectively.

Looking now the converse case, we show the distribution for the inner region of repeated tiles, again for short time scales, in Figure~\ref{subfig:rm_diff_hist_repeat_inner_short}. Here \qty{98}{\percent} of the component pairs are consistent with a standard normal distribution, with the remaining \qty{2}{\percent} fitted by a 0-mean component with a standard deviation of 10. We note that there is a similar result for the `repeat, inner, long' subset, with \qty{97}{\percent} fit by a standard normal distribution and the remaining fraction also fit by a 0-mean component with a standard deviation of 10.

Using this analysis alone we are unable to pinpoint the exact source of additional RM variance. Additionally, we are unable to distinguish between large variations in RM and underestimation of their uncertainties. However, as with our flux density analysis, we expect the a major source of systemic uncertainty to be in the characterisation of the primary beam towards the edge of the ASKAP tile. We are also reliant on the accuracy of the ionospheric reconstruction of \textsc{Frion} and \textsc{RMextract}~\citep{Mevius2018} in removing the ionospheric RM. Finally, our pencil-beam extraction method means that fluctuations in astrometric errors will result in different spatial components being extracted from our image cubes. This is broadly acceptable for a single RM grid experiment, but care must be taken for time-dependant variability analysis. Overall, we conclude that our component catalogue has a reliability in RM of \qty{98}{\percent} towards the centre of our observed tiles, which degrades to \qty{92}{\percent} towards the tile boundaries. Since we cannot strictly rule out the impact of astrophysical variation, we take these values as lower limits.

\section{Conclusion and outlook}\label{sec:conclusion}

We have presented the second data release (DR2) of SPICE-RACS. Using calibrated visibilities from the third low-band epoch of the Rapid ASKAP Continuum Survey, RACS-low3, we have imaged 1596 observations of 1493 unique fields in Stokes $I$, $Q$, and $U$. Our observations cover a total sky area of $\qty{3.5\pi}{\steradian}$ and a frequency range of \qtyrange{800}{1088}{\mega\hertz}. From these image cubes we extract polarisation spectra towards 5.5M radio components detected in total intensity. Around these components, we measure an average $rms$ sensitivity of $\sim\qty{200}{\micro\jansky\per PSF}$. We adopt a position dependent PSF, with an median major axis of $\ang{;;14.5}$, and extrema of $\ang{;;11.8}$ and \ang{;;75.9}. In total intensity we are able to fit a model spectrum to 3.1M unique components, and we find our flux scale is accurate within $\qty{15}{\percent}$.

In linear polarisation, our use of holographic primary beam models has allowed our residual polarisation leakage to be on the order of $\qty{0.1}{\percent}$. Using rotation measure (RM) synthesis we are able to measure the broad-band polarisation properties of our radio components. At a threshold of $8\sigma$ in linearly polarised signal-to-noise we detect 246509 components, making SPICE-RACS DR2 the largest polarisation catalogue ever produced by nearly an order of magnitude with an areal density of $6.7 \substack{+1.8\\-1.7}\,\unit{\perdegreesq}$. 

We further find that a threshold of $6\sigma$ in linear polarisation can provide a Gaussian-equivalent false detection rate of $5\sigma$. Applying such a threshold yields 342606 unique polarised components. Our derived RMs match very well existing catalogues, but our enhanced areal density has revealed a plethora of new magneto-ionic structures across the Southern sky. 

We find that $\sim\qty{1}{\percent}$ of our polarised components exhibit Faraday complexity on average, with a weak indication of enhanced complexity towards the Galactic plane. We advise careful interpretation of our complexity metrics, however, as we find they correlate with known instrumental effects. 

Finally, we are able to use the time domain information to estimate the reliability of our RM catalogue. Within the centre of region each tile we estimate a lower limit of \qty{98}{\percent}. Whereas towards the tile edges this degrades to a lower limit of \qty{92}{\percent}.

RACS was designed as a reference survey for the forthcoming deep ASKAP surveys, such as EMU and POSSUM. The latter of which will exceed the RM density achieved here by at least a factor of 5~\citep{Vanderwoude2024}. RACS has the unique advantages, however, of being both wider in sky area than planned for EMU/POSSUM \citep[nominally \qty{2\pi}{\steradian},][]{Gaensler2025} and broader in bandwidth. Combining RACS-low3 with RACS-mid and high would increase the sampled bandwidth by a factor of $\sim2.5$, and therefore reduce the $rms$ noise by a factor of $\sim1.6$. As briefly discussed in \citetalias{Duchesne2025}, using a combined RACS would improve the effective resolution in Faraday depth to $\delta\phi\approx\qty{30}{\radian\per\metre\squared}$. Additionally, the maximum Faraday depth scale would increase to $\phi_\text{max-scale}\approx\qty{110}{\radian\per\metre\squared}$, allowing for Faraday thick spectra to be detected and characterised. A polarisation survey with such properties will not be surpassed until the SKA surveys are executed. For now, SPICE-RACS DR2 provides wide coverage of the Southern sky with $\sim7$ RMs\,\unit{\perdegreesq} of excellent quality. This dataset will now enable a large subset of the POSSUM science case~\citep[see][]{Gaensler2025}, well ahead of the completion of the full ASKAP survey campaign. Furthermore, SPICE-RACS is placed as an ideal, wide-area reference for deeper observations with instruments such as ASKAP, MeerKAT, and the SKA.

\subsection{Data access}\label{sec:data_access}
All the calibrated visibilities for RACS-low3 are available on the CSIRO ASKAP Data Archive~\citet[CASDA][]{Huynh2020}\footnote{\url{https://data.csiro.au/domain/casda}}. All ASKAP Observatory RACS products can be found under project code AS110.

The data we have produced here, including image cubelets, spectra, and catalogues, are publicly available in a collection on the CSIRO Data Access Portal (DAP)\footnote{\url{https://data.csiro.au/collection/csiro:64891}}. 

We provide our primary polarised component catalogue in the \textsc{RMTable} schema, along with the initial Stokes $I$ catalogues used for source position references, all in FITS binary table format. The Stokes $I$ catalogues follow the same format presented in \citetalias{Duchesne2024} and \citetalias{Duchesne2025}. We stress that we do not consider these to be the primary release of the Stokes $I$ catalogues of RACS-low3; those data products are forthcoming (Galvin et al. in prep.).

For each total intensity source we produce image and weight cubelets in Stokes $I$, $Q$, and $U$ in FITS binary image format. We also provide image cubelets both with and without the application of ionospheric Faraday rotation correction. For each component we produce spectra in FITS binary table format. We produce a separate table for the frequency, FDF, and RMSF spectra. Users should take note that the Stokes $I$, $Q$, and $U$ frequency spectra have had background subtraction applied in-place. We do provide columns for the background for users wishing to undo this correction, along with our estimated $rms$ noise. Due to the large data volume, the cubelets and spectra are collected into tar files for each SBID. 

We additionally provide a number of sky maps in HEALPix FITS format. We produce these maps via nearest-neighbour and inverse-distance weighted interpolation onto a $N_\text{side}=512$ grid for the following catalogue columns: \texttt{rm}, \texttt{nn\_rm\_med}, \texttt{nn\_rm\_mad\_std}, \texttt{nn\_rm\_se}, \texttt{nn\_rm\_wmean}, \texttt{nn\_rm\_wmean\_err}, \texttt{nn\_rm\_sep\_max\_deg}, \texttt{nn\_rm\_sep\_min\_deg}. We provide interpolation routines publicly on GitHub\footnote{\url{https://github.com/AlecThomson/sphinterp}} for users wishing to produce additional sky maps.

Further access details and data layout are described in detail in the DAP collection.

\nocite{
    Taylor2009,
    Thomson2023,
    Schnitzeler2019,
    OSullivan2023,
    Farnes2014,
    Mao2010,
    2021ApJS..253...48V,
    Betti2019,
    Tabara1980,
    Riseley2020,
    Mao2012,
    2011ApJ...728...97V,
    2012ApJ...755...21M,
    2020MNRAS.497.3097M,
    2017MNRAS.467.1776K,
    2003A&A...406..579K,
    Manchester2005,
    2007ApJ...663..258B,
    2009ApJ...707..114F,
    Vanderwoude2024,
    2015ApJ...815...49A,
    2017MNRAS.469.4034O,
    2001ApJ...547L.111C,
    2024A&A...686A.104R,
    Riseley2018,
    1996ApJ...458..194M,
    Mao2008,
    2003ApJS..145..213B,
    Loi2025,
    2016ApJ...829..133K,
    1988Ap&SS.141..303B,
    1981ApJS...45...97S,
    1992ApJ...386..143C,
    2019ApJ...887L...7S,
    Livingston2021,
    Law2011,
    Livingston2022,
    Ma2019,
    2005MNRAS.360.1305R,
    Gaensler2001,
    2018A&A...613A..58V,
    Heald2009,
    Taylor2024,
    2018MNRAS.475.1736V,
    Taylor2024a,
    1995ApJ...445..624O%
}

\begin{acknowledgement}
The authors would like to thank the anonymous referee for their constructive review of this work.

This scientific work uses data obtained from Inyarrimanha Ilgari Bundara, the CSIRO Murchison Radio-astronomy Observatory. We acknowledge the Wajarri Yamaji People as the Traditional Owners and native title holders of the Observatory site. CSIRO's ASKAP radio telescope is part of the Australia Telescope National Facility (\url{https://ror.org/05qajvd42}). Operation of ASKAP is funded by the Australian Government with support from the National Collaborative Research Infrastructure Strategy. ASKAP uses the resources of the Pawsey Supercomputing Research Centre. Establishment of ASKAP, Inyarrimanha Ilgari Bundara, the CSIRO Murchison Radio-astronomy Observatory and the Pawsey Supercomputing Research Centre are initiatives of the Australian Government, with support from the Government of Western Australia and the Science and Industry Endowment Fund.

This project was supported by resources and expertise provided by CSIRO IMT Scientific Computing.

This work was made possible by large number of open source software packages and libraries. Some of the results in this paper have been derived using the \textsc{healpy} and HEALPix packages~\citep{Gorski2005}. This research made use of: NASA's Astrophysics Data System; \textsc{spectralcube}, a library for astronomical spectral data cubes; \textsc{Astropy}\footnote{\url{http://www.astropy.org}}, a community-developed core Python package and an ecosystem of tools and resources for astronomy \citep{astropy_2013, astropy_2018, astropy_2022}; \textsc{SciPy} \citep{2020SciPy-NMeth}; \textsc{ds9}, a tool for data visualization supported by the Chandra X-ray Science Center (CXC) and the High Energy Astrophysics Science Archive Center (HEASARC) with support from the JWST Mission office at the Space Telescope Science Institute for 3D visualization; \textsc{NumPy} \citep{Harris2020}; \textsc{matplotlib}, a Python library for publication-quality graphics \citep{Hunter:2007}. This work made use of the \textsc{IPython} package \citep{PER-GRA:2007}; \textsc{pandas} \citep{McKinney_2010, McKinney_2011}; TOPCAT, an interactive graphical viewer and editor for tabular data \citep{2005ASPC..347...29T}

\end{acknowledgement}

\paragraph{Funding Statement}

SPO acknowledges support from the Comunidad de Madrid Atracción de Talento program via grant 2022-T1/TIC-23797.  SPO, SM acknowledge support from the grant PID2023-146372OB-I00 funded by MICIU/AEI/10.13039/501100011033 and by ERDF, EU. SH  acknowledges funding by the European Union (ERC, ISM-FLOW, 101055318). Views and opinions expressed are, however, those of the author(s) only and do not necessarily reflect those of the European Union or the European Research Council. Neither the European Union nor the granting authority can be held responsible for them. 

\paragraph{Competing Interests}

None.

\paragraph{Data Availability Statement}

The data underlying and produced by this work are publicly available. Details of data access are provided in \S\ref{sec:data_access}. The scripts used to produce the analysis and diagrams in this work are available upon reasonable request to the authors.

\paragraph{Generative AI and Large Language Models}

No large language models (LLMs) or generative AI tools were used to draft this manuscript, nor in the underlying scientific research. 

The in-line autocompletion of GitHub Copilot\footnote{\url{https://docs.github.com/en/copilot}} was used to assist in the development of software used in this paper.

%% file: tables/new_pulsars.tex
\begin{table}
\caption{Known pulsars with new RMs in SPICE-RACS DR2.}
\label{tab:pulsar}
\begin{tabular}{lcc}
\toprule
\verb|cat_id| & Pulsar JName & RM \\
\midrule
\verb|RACS_1833-73_7276| & J1846-7403 & $28\pm3$ \\
\verb|RACS_1733+00_8209| & J1726-00 & $42\pm3$ \\
\verb|RACS_1326+32_3837| & J1327+3423 & $0\pm3$ \\
\verb|RACS_1817-23_3937| & J1824-2452 & $79\pm3$ \\
\verb|RACS_1724-28_9794| & J1712-2715 & $29\pm3$ \\
\verb|RACS_1800+14_5541| & J1803+1358 & $74\pm3$ \\
\bottomrule
\end{tabular}
\end{table}

%% file: appendix.tex

\onecolumn
\section{Example Configuration File}\label{sec:config_file}

\begin{center}
\captionof{lstlisting}{An example Arrakis configuration YAML file.}
\label{lst:config}
\begin{minipage}[t]{0.49\textwidth}
\begin{lstlisting}[language=yaml, firstline=1, lastline=41]
# 56228_rm.yaml
# pipeline options
skip_cat: false
skip_cleanup: false
skip_cutout: false
skip_fix_ms: false
skip_frion: false
skip_imager: false
skip_linmos: false
skip_rmclean: false
skip_rmsynth: false
skip_validate: false
dask_config: /scratch3/projects/spiceracs/processing/proc_scripts/rm_petrichor.yaml
imager_dask_config: /scratch3/projects/spiceracs/processing/proc_scripts/petrichor_big.yaml
host: mueller.it.csiro.au
username: admin
password: "********"
database: true
# imager options
gridder: wgridder
size: 6144
absmem: 128
robust: -0.5
auto_mask: 5
auto_threshold: 0.5
multiscale: false
multiscale_scale_bias: 0.6
multiscale_scales: 0,2,4,8,16,32,64,128
nchan: 72
niter: 500000
nmiter: 15
pols: IQU
psf_cutoff: 19.0
purge: true
temp_dir_images: $MEMDIR
temp_dir_wsclean: $MEMDIR
data_column: CORRECTED_DATA
mgain: 0.6
minuv: 200
ms_glob_pattern: "scienceData.RACS_1800-32.SB56228.RACS_1800-32.beam*_averaged_cal.leakage.ms"
scale: 2.5
\end{lstlisting}
\end{minipage}%
\hfill
\begin{minipage}[t]{0.49\textwidth}
\begin{lstlisting}[language=yaml, firstline=42, firstnumber=42]
imager_only: false
local_rms: true
local_rms_window: 60
local_wsclean: /datasets/work/sa-spiceracs/work/containers/wsclean_force_mask.sif
disable_pol_force_mask_rounds: true
disable_pol_local_rms: true
# linmos options
holofile: /scratch3/projects/spiceracs/RACS_Low3_Holography/akpb.iquv.closepack36.54.943MHz.SB55219.cube.fits
yanda_image: /datasets/work/sa-spiceracs/work/containers/askapsoft_1.16.1-openmpi5.sif
# rmsynth/clean options
cutoff: -8.0
window: -5.0
weight_type: variance
debug: false
dimension: 1d
dryrun: false
epoch: 9
fit_function: log
fit_rmsf: true
force_mask_rounds: 8
gain: 0.1
max_iter: 10000
n_samples: 100.0
no_stokes_i: false
not_rmsf: false
pad: 7.0
save_plots: true
show_plots: false
poly_ord: -2
rm_verbose: false
# frion options
ionex_formatter: cddis.gsfc.nasa.gov
ionex_prefix: codg
ionex_server: file:///datasets/work/sa-spiceracs/work/data/IONEX/gdc.cddis.eosdis.nasa.gov
# Catalogue and validation options
catfile: RACS_1800-32_SB56228_polcat.fits
leakage_bins: 8
leakage_degree: 2

\end{lstlisting}
\end{minipage}
\end{center}
\twocolumn

\onecolumn
\section{Conversion of ASKAP correlations to the IAU frame}\label{sec:vis_convert}
In an ASKAP antenna's frame of reference, the instrumental linear `horizontal' ($H$) and `vertical' ($V$) feeds are inclined at \ang{-45} and \ang{+45} to the antenna's reference direction, respectively~\citep{Reynolds2011aces}. The angle `$\mathcal{P}$' is measured counter-clockwise from the antenna's polarisation reference to the North Celestial Pole \citep[denoted as `$X$', as per the International Astronomical Union (IAU) definition,][]{IAU_1973}.

During normal observations, the antenna is rotated continuously about its roll axis to keep the polarisation vectors fixed relative to celestial coordinates such that the angle $\mathcal{P}$ is kept fixed to a value specified in the observing parset. In this way, the $H$ and $V$ feeds are inclined at a fixed offset to celestial North. \textsc{ASKAPsoft} forms IAU-conventional Stokes parameters during imaging using~\citep{AKVET2023}
\begin{equation}\label{eqn:askap_stokes}
\begin{pmatrix}
I\\
Q\\
U\\
V
\end{pmatrix} = 
\begin{pmatrix}
1 & 0 & 0 & 1 \\
\sin 2\mathcal{P} & \cos 2\mathcal{P} & \cos 2\mathcal{P} & -\sin 2\mathcal{P}\\
-\cos 2\mathcal{P} & \sin 2\mathcal{P} & \sin 2\mathcal{P} & \cos 2\mathcal{P}\\
0 & -j & j & 0
\end{pmatrix}
\begin{pmatrix}
HH\\
HV\\
VH\\
VV
\end{pmatrix}.
\end{equation}

In the ASKAP observation parset, which is parsed by the Telescope Operating System (TOS), the offset of the reference direction is given by:
\begin{lstlisting}[mathescape=true]
    pol_axis = [pa_fixed, $\mathcal{P}$]
\end{lstlisting}
After an observation, this value can be retrieved directly from the online Observation Management Portal (OMP)\footnote{\url{https://apps.atnf.csiro.au/OMP}}. The value of $\mathcal{P}$ also recorded indirectly in the visibilities MeasurementSet \texttt{FEED} table. Inspecting the \texttt{RECEPTOR\_ANGLE} column of the \texttt{FEED} table you will find a $(N,2)$ array, where $N$ is the number of antennas by the number of beams (typically $36\times36=1296$). The first column in this array corresponds to the instrumental $H$ feed, and the second column to the $V$ feed. The \texttt{RECEPTOR\_ANGLE} value ($\mathcal{R}$) is calculated as reference mounting angle of the feed minus the polarisation reference angle $\mathcal{P}$. The values of the mounting angles could vary in theory, but in practice they are fixed at $+45\degr$ for all antennas. So, for the $H$ feed:
\begin{equation}
    \mathcal{R} = \frac{\pi}{4} - \mathcal{P}
\end{equation}
and for the $V$ feed:
\begin{equation}
    \mathcal{R} = \frac{\pi}{4} - \mathcal{P} + \frac{\pi}{2}.
\end{equation}
This implies that the $H$ and $V$ feeds are perfectly orthogonal, and therefore any minor deviations are treated as leakage.

In contrast to \textsc{ASKAPsoft}, both \textsc{CASA} and \textsc{WSClean} apply the following equation to form Stokes parameters:

\begin{equation}\label{eqn:casa_stokes}
    \begin{pmatrix}
    I\\
    Q\\
    U\\
    V
    \end{pmatrix} = 
    \begin{pmatrix}
    1/2 & 0 & 0 & 1/2 \\
    1/2& 0& 0& -1/2\\
    0& 1/2& 1/2& 0\\
    0& -j/2& j/2& 0
    \end{pmatrix}
    \begin{pmatrix}
    XX\\
    XY\\
    YX\\
    YY
    \end{pmatrix}.
\end{equation}
Here $X$ and $Y$ are aligned with celestial North and East, respectively, following the IAU convention. Equating Equations~\ref{eqn:askap_stokes} and \ref{eqn:casa_stokes} we can solve for the sky frame correlations in terms of the instrumental correlations:
\begin{equation}\label{eqn:askap_to_casa}
    \begin{pmatrix}
        XX\\XY\\YX\\YY
    \end{pmatrix} = 
    \begin{pmatrix}
        \sin{\left(2\mathcal{P} \right)} + 1 & \cos{\left(2\mathcal{P} \right)} & \cos{\left(2\mathcal{P} \right)} & 1 - \sin{\left(2\mathcal{P} \right)}\\
        - \cos{\left(2\mathcal{P} \right)} & \sin{\left(2\mathcal{P} \right)} + 1 & \sin{\left(2\mathcal{P} \right)} - 1 & \cos{\left(2\mathcal{P} \right)}\\
        - \cos{\left(2\mathcal{P} \right)} & \sin{\left(2\mathcal{P} \right)} - 1 & \sin{\left(2\mathcal{P} \right)} + 1 & \cos{\left(2\mathcal{P} \right)}\\
        1 - \sin{\left(2\mathcal{P} \right)} & - \cos{\left(2\mathcal{P} \right)} & - \cos{\left(2\mathcal{P} \right)} & \sin{\left(2\mathcal{P} \right)} + 1
    \end{pmatrix}
    \begin{pmatrix}
        HH\\HV\\VH\\VV
    \end{pmatrix}.
\end{equation}
Note that this matrix both transforms the instrumental correlations to the sky frame, and also changes the `factor of 2' convention used by \textsc{ASKAPsoft} from $I = XX + YY$ to $I = (XX + YY)/2$. This correction matrix (Equation~\ref{eqn:askap_to_casa}) is implemented in \textsc{FixMS} within the \verb|fix_ms_corrs| routine.

\twocolumn

\onecolumn
\section{Catalogue sample}
\input{tables/cat.tex}
\twocolumn

\begin{landscape}
\section{RM Grid in Galactic coordinates}
\vfill
\centering
\includegraphics{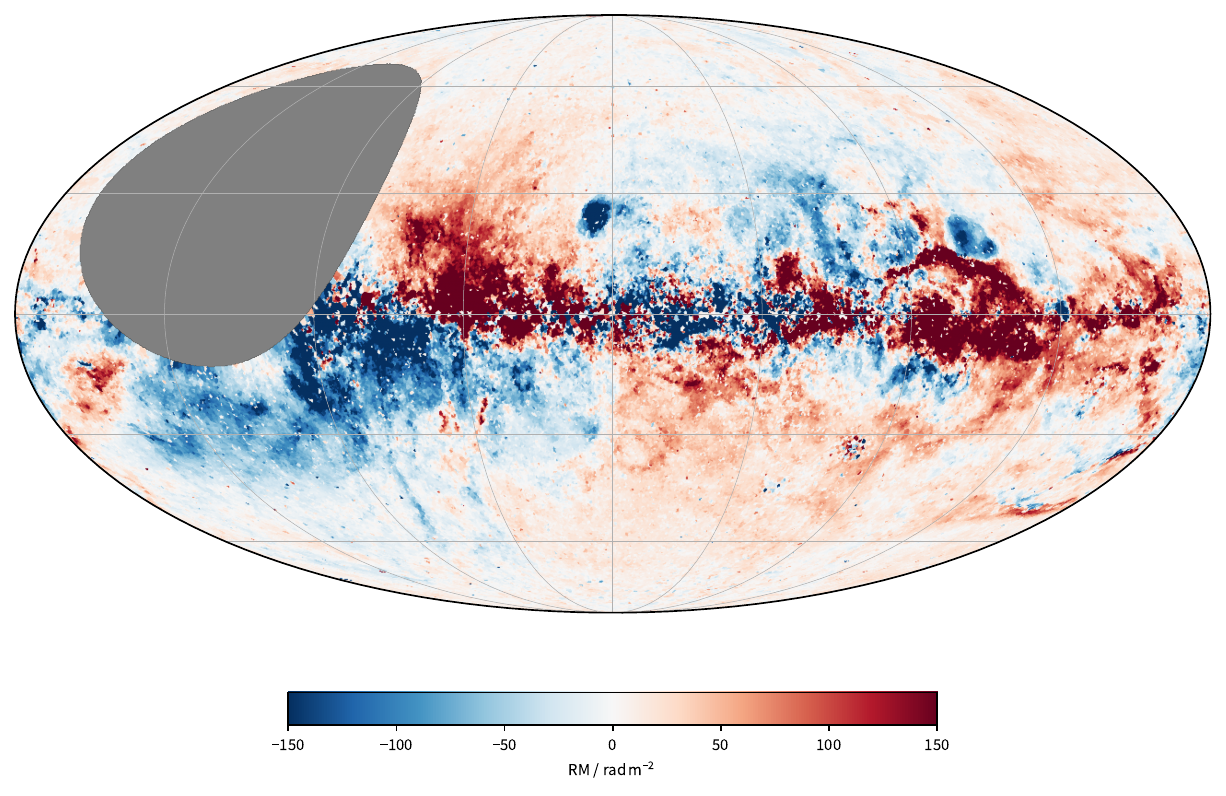}
\captionof{figure}{Rotation measures (RM) across the survey area in Galactic coordinates using nearest-neighbour interpolation. The data is the same as in Figure~\ref{fig:rm_non_linear}, but here we use both a diverging colour map and a linear scale to highlight different features of the RM sky.}
\label{fig:rm_linear}
\vfill
\end{landscape}





\onecolumn

%% file: tables/cat.tex
\begin{longtable}{l|ccc}
\caption {The first two rows of the SPICE-RACS DR2 catalogue. We have transposed the table for readability. We define all column names in \S\ref{sec:catalogue}.}\label{tab:catalogue}\\
\toprule
\endfirsthead
\caption*{\textbf{Table \ref{tab:catalogue} Continued}}\\
\endhead
\endfoot
\bottomrule
\endlastfoot
Column & Units & Row 1 & Row 2 \\
\midrule
\texttt{source\_id} & $\mathrm{---}$ & \texttt{RACS\_1909+41\_2824} & \texttt{RACS\_1909+41\_2868} \\
\texttt{cat\_id} & $\mathrm{---}$ & \texttt{RACS\_1909+41\_3540} & \texttt{RACS\_1909+41\_3591} \\
\texttt{ra} & $\mathrm{{}^{\circ}}$ & \texttt{287.93490403998845} & \texttt{287.88809793513957} \\
\texttt{ra\_err} & $\mathrm{{}^{\circ}}$ & \texttt{7.14593897618979e-05} & \texttt{0.00033589006768102554} \\
\texttt{dec} & $\mathrm{{}^{\circ}}$ & \texttt{39.966451278970254} & \texttt{42.13892402515704} \\
\texttt{dec\_err} & $\mathrm{{}^{\circ}}$ & \texttt{0.00031445457068518144} & \texttt{0.0019090434271681414} \\
\texttt{total\_I\_flux} & $\mathrm{Jy}$ & \texttt{0.012487492094573805} & \texttt{0.0017804733995072483} \\
\texttt{total\_I\_flux\_err} & $\mathrm{Jy}$ & \texttt{0.0009577063477698441} & \texttt{0.0005500349516795522} \\
\texttt{peak\_I\_flux} & $\mathrm{Jy\,PSF^{-1}}$ & \texttt{0.007299706380568457} & \texttt{0.0012183822446276006} \\
\texttt{peak\_I\_flux\_err} & $\mathrm{Jy\,PSF^{-1}}$ & \texttt{0.0004918049612982018} & \texttt{0.00023188433297717789} \\
\texttt{maj\_axis} & $\mathrm{{}^{\prime\prime}}$ & \texttt{62.80485394968595} & \texttt{60.4519943020728} \\
\texttt{maj\_axis\_err} & $\mathrm{{}^{\prime\prime}}$ & \texttt{2.671589275914708} & \texttt{16.18535883573754} \\
\texttt{min\_axis} & $\mathrm{{}^{\prime\prime}}$ & \texttt{26.251234867943836} & \texttt{23.284905698298612} \\
\texttt{min\_axis\_err} & $\mathrm{{}^{\prime\prime}}$ & \texttt{0.5795762811255014} & \texttt{2.837115330977609} \\
\texttt{pa} & $\mathrm{{}^{\circ}}$ & \texttt{4.258349995290576} & \texttt{1.1696668314083982} \\
\texttt{pa\_err} & $\mathrm{{}^{\circ}}$ & \texttt{2.5700600492783057} & \texttt{11.980761385452757} \\
\texttt{dc\_maj\_axis} & $\mathrm{{}^{\prime\prime}}$ & \texttt{28.50262542699576} & \texttt{23.5379148031914} \\
\texttt{dc\_maj\_axis\_err} & $\mathrm{{}^{\prime\prime}}$ & \texttt{2.671589275914708} & \texttt{16.18535883573754} \\
\texttt{dc\_min\_axis} & $\mathrm{{}^{\prime\prime}}$ & \texttt{19.64071334432083} & \texttt{14.514961554447941} \\
\texttt{dc\_min\_axis\_err} & $\mathrm{{}^{\prime\prime}}$ & \texttt{0.5795762811255014} & \texttt{2.837115330977609} \\
\texttt{dc\_pa} & $\mathrm{{}^{\circ}}$ & \texttt{11.378361175412289} & \texttt{163.3607343868265} \\
\texttt{dc\_pa\_err} & $\mathrm{{}^{\circ}}$ & \texttt{2.5700600492783057} & \texttt{11.980761385452757} \\
\texttt{stokesI\_err} & $\mathrm{Jy\,PSF^{-1}}$ & \texttt{0.0002401885711286148} & \texttt{0.0001969266778798084} \\
\texttt{stokesQ\_err} & $\mathrm{Jy\,PSF^{-1}}$ & \texttt{0.00021148871881585497} & \texttt{0.00017491710793735734} \\
\texttt{stokesU\_err} & $\mathrm{Jy\,PSF^{-1}}$ & \texttt{0.0001932935038995412} & \texttt{0.00017290450718805746} \\
\texttt{stokesI\_bkg} & $\mathrm{Jy\,PSF^{-1}}$ & \texttt{0.0005461489126901142} & \texttt{3.556922039630687e-05} \\
\texttt{stokesQ\_bkg} & $\mathrm{Jy\,PSF^{-1}}$ & \texttt{4.712333954002032e-05} & \texttt{-1.287478841681554e-05} \\
\texttt{stokesU\_bkg} & $\mathrm{Jy\,PSF^{-1}}$ & \texttt{5.286082291730862e-05} & \texttt{2.0961897032773573e-05} \\
\texttt{beamdist} & $\mathrm{{}^{\circ}}$ & \texttt{1.9678118031665726} & \texttt{0.5049121474410881} \\
\texttt{N\_Gaus} & $\mathrm{---}$ & \texttt{4} & \texttt{1} \\
\texttt{tile\_id} & $\mathrm{---}$ & \texttt{RACS\_1909+41} & \texttt{RACS\_1909+41} \\
\texttt{sbid} & $\mathrm{---}$ & \texttt{55730} & \texttt{55730} \\
\texttt{start\_time} & $\mathrm{d}$ & \texttt{60301.21156267109} & \texttt{60301.21156267109} \\
\texttt{separation\_tile\_centre} & $\mathrm{{}^{\circ}}$ & \texttt{1.9678118031665726} & \texttt{0.5049121474410881} \\
\texttt{s\_code} & $\mathrm{---}$ & \texttt{M} & \texttt{S} \\
\texttt{rm} & $\mathrm{rad\,m^{-2}}$ & \texttt{35.17746095420923} & \texttt{629.528869358685} \\
\texttt{rm\_err} & $\mathrm{rad\,m^{-2}}$ & \texttt{10.001452603866586} & \texttt{10.845458164225702} \\
\texttt{polint} & $\mathrm{Jy\,PSF^{-1}}$ & \texttt{0.0006033722312673457} & \texttt{0.0004863881822067386} \\
\texttt{polint\_err} & $\mathrm{Jy\,PSF^{-1}}$ & \texttt{0.00019193085794653024} & \texttt{0.00017087770190027482} \\
\texttt{stokesQ} & $\mathrm{Jy\,PSF^{-1}}$ & \texttt{-0.00048342346333231286} & \texttt{0.0004171676218069758} \\
\texttt{stokesU} & $\mathrm{Jy\,PSF^{-1}}$ & \texttt{0.0003610535255632517} & \texttt{-0.0002500832254381558} \\
\texttt{polangle} & $\mathrm{{}^{\circ}}$ & \texttt{71.62256128722069} & \texttt{164.52909909356705} \\
\texttt{polangle\_err} & $\mathrm{{}^{\circ}}$ & \texttt{9.112805950286251} & \texttt{10.064565186761556} \\
\texttt{derot\_polangle} & $\mathrm{{}^{\circ}}$ & \texttt{49.82258622976974} & \texttt{100.34863920069847} \\
\texttt{derot\_polangle\_err} & $\mathrm{{}^{\circ}}$ & \texttt{51.08558731613811} & \texttt{57.21925884202857} \\
\texttt{fracpol} & $\mathrm{---}$ & \texttt{0.18192398180298186} & \texttt{0.48996458276733407} \\
\texttt{reffreq\_pol} & $\mathrm{Hz}$ & \texttt{947444349.12514} & \texttt{940592296.3550439} \\
\texttt{reffreq\_beam} & $\mathrm{Hz}$ & \texttt{947444349.12514} & \texttt{940592296.3550439} \\
\texttt{reffreq\_I} & $\mathrm{Hz}$ & \texttt{947444349.12514} & \texttt{940592296.3550439} \\
\texttt{fdf\_noise\_th} & $\mathrm{Jy\,PSF^{-1}}$ & \texttt{0.00019193085794653024} & \texttt{0.00017087770190027482} \\
\texttt{rmsf\_fwhm} & $\mathrm{rad\,m^{-2}}$ & \texttt{62.88304901123047} & \texttt{61.74126434326172} \\
\texttt{refwave\_sq\_pol} & $\mathrm{m^{2}}$ & \texttt{0.10012303047623769} & \texttt{0.10158710191880463} \\
\texttt{stokesI} & $\mathrm{Jy\,PSF^{-1}}$ & \texttt{0.003316617349915564} & \texttt{0.0009927006564662772} \\
\texttt{stokesI\_fit\_flag\_is\_negative} & $\mathrm{---}$ & \texttt{False} & \texttt{False} \\
\texttt{stokesI\_fit\_flag\_is\_close\_to\_zero} & $\mathrm{---}$ & \texttt{False} & \texttt{True} \\
\texttt{stokesI\_fit\_flag\_is\_not\_finite} & $\mathrm{---}$ & \texttt{False} & \texttt{True} \\
\texttt{stokesI\_fit\_flag\_is\_not\_normal} & $\mathrm{---}$ & \texttt{False} & \texttt{True} \\
\texttt{stokesI\_chi2\_red} & $\mathrm{---}$ & \texttt{0.618149754964099} & \texttt{1.446236397104928} \\
\texttt{snr\_polint} & $\mathrm{---}$ & \texttt{3.1436957960946454} & \texttt{2.846411069424362} \\
\texttt{minfreq} & $\mathrm{Hz}$ & \texttt{803490740.7407} & \texttt{803490740.7407} \\
\texttt{maxfreq} & $\mathrm{Hz}$ & \texttt{1083490740.7407} & \texttt{1083490740.7407} \\
\texttt{channelwidth} & $\mathrm{Hz}$ & \texttt{8000000.0} & \texttt{8000000.0} \\
\texttt{Nchan} & $\mathrm{---}$ & \texttt{36} & \texttt{36} \\
\texttt{rm\_width} & $\mathrm{rad\,m^{-2}}$ & \texttt{nan} & \texttt{nan} \\
\texttt{stokesI\_model\_coef} & $\mathrm{---}$ & \makecell{\texttt{0.0},\\\texttt{0.0},\\\texttt{0.0},\\\texttt{0.0},\\\texttt{0.0},\\\texttt{0.0033166173}} & \makecell{\texttt{0.0},\\\texttt{0.0},\\\texttt{0.0},\\\texttt{0.0},\\\texttt{0.0},\\\texttt{0.0009927007}} \\
\texttt{stokesI\_model\_coef\_err} & $\mathrm{---}$ & \makecell{\texttt{0.0},\\\texttt{0.0},\\\texttt{0.0},\\\texttt{0.0},\\\texttt{0.0},\\\texttt{0.00022816521}} & \makecell{\texttt{0.0},\\\texttt{0.0},\\\texttt{0.0},\\\texttt{0.0},\\\texttt{0.0},\\\texttt{0.00018642988}} \\
\texttt{stokesI\_model\_order} & $\mathrm{---}$ & \texttt{0.0} & \texttt{0.0} \\
\texttt{noise\_chan} & $\mathrm{Jy\,PSF^{-1}}$ & \texttt{0.0011433771724114195} & \texttt{0.0010356951534049585} \\
\texttt{fdf\_noise\_mad} & $\mathrm{Jy\,PSF^{-1}}$ & \texttt{0.00023191826767288148} & \texttt{0.00013928351108916104} \\
\texttt{fdf\_noise\_rms} & $\mathrm{Jy\,PSF^{-1}}$ & \texttt{0.00019193085794653024} & \texttt{0.00017087770190027482} \\
\texttt{beam\_maj} & $\mathrm{{}^{\circ}}$ & \texttt{0.00955555555555555} & \texttt{0.00944444444444444} \\
\texttt{beam\_min} & $\mathrm{{}^{\circ}}$ & \texttt{0.00280555555555556} & \texttt{0.00275} \\
\texttt{beam\_pa} & $\mathrm{{}^{\circ}}$ & \texttt{0.08} & \texttt{0.27} \\
\texttt{l\_tile\_centre} & $\mathrm{{}^{\circ}}$ & \texttt{0.4807934526606305} & \texttt{0.43038119000506575} \\
\texttt{m\_tile\_centre} & $\mathrm{{}^{\circ}}$ & \texttt{-1.908172148565309} & \texttt{0.26402330943194136} \\
\texttt{sigma\_add} & $\mathrm{---}$ & \texttt{0.12326492979643032} & \texttt{0.09253382365571675} \\
\texttt{sigma\_add\_err} & $\mathrm{---}$ & \texttt{1.8126992180330121} & \texttt{0.5751948953587795} \\
\texttt{snr\_flag} & $\mathrm{---}$ & \texttt{True} & \texttt{True} \\
\texttt{leakage\_flag} & $\mathrm{---}$ & \texttt{False} & \texttt{False} \\
\texttt{channel\_flag} & $\mathrm{---}$ & \texttt{False} & \texttt{False} \\
\texttt{stokesI\_fit\_flag} & $\mathrm{---}$ & \texttt{False} & \texttt{True} \\
\texttt{complex\_sigma\_add\_flag} & $\mathrm{---}$ & \texttt{False} & \texttt{False} \\
\texttt{complex\_M2\_CC\_flag} & $\mathrm{---}$ & \texttt{False} & \texttt{False} \\
\texttt{local\_rm\_flag} & $\mathrm{---}$ & \texttt{False} & \texttt{False} \\
\texttt{is\_blended\_flag} & $\mathrm{---}$ & \texttt{False} & \texttt{False} \\
\texttt{blend\_ratio} & $\mathrm{---}$ & \texttt{1.0} & \texttt{nan} \\
\texttt{N\_blended} & $\mathrm{---}$ & \texttt{0} & \texttt{0} \\
\texttt{spectral\_index} & $\mathrm{---}$ & \texttt{0.0} & \texttt{0.0} \\
\texttt{spectral\_index\_err} & $\mathrm{---}$ & \texttt{0.0} & \texttt{0.0} \\
\texttt{int\_time} & $\mathrm{s}$ & \texttt{905.74848} & \texttt{905.74848} \\
\texttt{epoch} & $\mathrm{d}$ & \texttt{60301.21680427109} & \texttt{60301.21680427109} \\
\texttt{l} & $\mathrm{{}^{\circ}}$ & \texttt{71.21721173380412} & \texttt{73.24650342387741} \\
\texttt{b} & $\mathrm{{}^{\circ}}$ & \texttt{13.44738280782637} & \texttt{14.36434833683301} \\
\texttt{pos\_err} & $\mathrm{{}^{\prime\prime}}$ & \texttt{0.007682104037589461} & \texttt{0.007682104037589461} \\
\texttt{rm\_method} & $\mathrm{---}$ & \makecell{\texttt{RM Synthesis -},\\\texttt{ Fractional polarization}} & \makecell{\texttt{RM Synthesis -},\\\texttt{ Fractional polarization}} \\
\texttt{telescope} & $\mathrm{---}$ & \texttt{ASKAP} & \texttt{ASKAP} \\
\texttt{pol\_bias} & $\mathrm{---}$ & \texttt{2012PASA...29..214G} & \texttt{2012PASA...29..214G} \\
\texttt{ionosphere} & $\mathrm{---}$ & \texttt{FRion} & \texttt{FRion} \\
\texttt{flux\_type} & $\mathrm{---}$ & \texttt{Peak} & \texttt{Peak} \\
\texttt{aperture} & $\mathrm{{}^{\circ}}$ & \texttt{0.0} & \texttt{0.0} \\
\texttt{leakage} & $\mathrm{---}$ & \texttt{0.0012007311664392227} & \texttt{0.0011585383757299272} \\
\texttt{complex\_flag} & $\mathrm{---}$ & \texttt{False} & \texttt{False} \\
\texttt{rm\_width\_err} & $\mathrm{rad\,m^{-2}}$ & \texttt{nan} & \texttt{nan} \\
\texttt{complex\_test} & $\mathrm{---}$ & \texttt{s} & \texttt{s} \\
\texttt{Ncomp} & $\mathrm{---}$ & \texttt{1} & \texttt{1} \\
\texttt{fracpol\_err} & $\mathrm{---}$ & \texttt{nan} & \texttt{nan} \\
\texttt{stokesV} & $\mathrm{Jy}$ & \texttt{nan} & \texttt{nan} \\
\texttt{stokesV\_err} & $\mathrm{Jy}$ & \texttt{nan} & \texttt{nan} \\
\texttt{obs\_interval} & $\mathrm{d}$ & \texttt{nan} & \texttt{nan} \\
\texttt{catalog\_name} & $\mathrm{---}$ & \texttt{SPICE-RACS DR2} & \texttt{SPICE-RACS DR2} \\
\texttt{dataref} & $\mathrm{---}$ & \makecell{\texttt{https://},\\\texttt{data.csiro.au/},\\\texttt{collection/},\\\texttt{csiro:64891/}} & \makecell{\texttt{https://},\\\texttt{data.csiro.au/},\\\texttt{collection/},\\\texttt{csiro:64891/}} \\
\texttt{type} & $\mathrm{---}$ & $\mathrm{---}$ & $\mathrm{---}$ \\
\texttt{notes} & $\mathrm{---}$ & $\mathrm{---}$ & $\mathrm{---}$ \\
\texttt{goodI\_flag} & $\mathrm{---}$ & \texttt{True} & \texttt{False} \\
\texttt{goodRM\_flag} & $\mathrm{---}$ & \texttt{False} & \texttt{False} \\
\texttt{nn\_rm\_count} & $\mathrm{---}$ & \texttt{35} & \texttt{36} \\
\texttt{nn\_rm\_med} & $\mathrm{rad\,m^{-2}}$ & \texttt{33.38595667955203} & \texttt{1.4273758902960954} \\
\texttt{nn\_rm\_mean} & $\mathrm{rad\,m^{-2}}$ & \texttt{27.50011728298653} & \texttt{1.7293585863622465} \\
\texttt{nn\_rm\_mad\_std} & $\mathrm{rad\,m^{-2}}$ & \texttt{15.77296678853631} & \texttt{16.07999156024331} \\
\texttt{nn\_rm\_std} & $\mathrm{rad\,m^{-2}}$ & \texttt{18.409056254811176} & \texttt{14.89286244594474} \\
\texttt{nn\_rm\_se} & $\mathrm{rad\,m^{-2}}$ & \texttt{3.1116984438580553} & \texttt{2.48214374099079} \\
\texttt{nn\_rm\_wmean} & $\mathrm{rad\,m^{-2}}$ & \texttt{33.27748355776515} & \texttt{-0.4422982025423753} \\
\texttt{nn\_rm\_wmean\_err} & $\mathrm{rad\,m^{-2}}$ & \texttt{0.2091919578794649} & \texttt{0.2512882043093384} \\
\texttt{nn\_rm\_sep\_max\_deg} & $\mathrm{{}^{\circ}}$ & \texttt{1.5548291256730578} & \texttt{1.6644765034645443} \\
\texttt{nn\_rm\_sep\_min\_deg} & $\mathrm{{}^{\circ}}$ & \texttt{0.25868603539792484} & \texttt{0.47272277552332304} \\
\end{longtable}